\documentclass[twocolumn,showpacs,amssymb,prb]{revtex4}

\newcommand{\be}{\begin{equation}}
\newcommand{\ee}{\end{equation}}
\usepackage{graphicx}
\usepackage{epsfig}

\begin{document}

\title{Aging Dynamics of the Heisenberg Spin Glass}

\author{L.~Berthier}
\email{berthier@thphys.ox.ac.uk}
\homepage{http://www-thphys.physics.ox.ac.uk/users/LudovicBerthier}
\affiliation{Theoretical Physics,
1 Keble Road, Oxford OX1 3NP, UK and \\
Laboratoire des Verres, Universit\'e Montpellier II, 34095
Montpellier, France}

\author{A.~P.~Young}
\email{peter@bartok.ucsc.edu}
\homepage{http://bartok.ucsc.edu/peter}
\affiliation{Department of Physics,
University of California,
Santa Cruz, California 95064}

\date{\today}

\begin{abstract}
We numerically study the non-equilibrium dynamics of the
three dimensional Heisenberg Edwards-Anderson spin glass 
following a sudden quench to its low temperature phase. 
The subsequent aging behavior of the system 
is analyzed in detail, and the scaling behavior of various space-time 
correlation functions is investigated for both spin and chiral
degrees of freedom.
We carefully compare our results with those obtained from simulations of
the more studied Ising version of the model, and
with experiments on real spin glasses in which the spins
have vectorial character. 
Finally, the present dynamical study offers new perspectives into 
the possibility of spin-chirality decoupling at 
low temperature in vectorial spin glasses.
\end{abstract}

\maketitle

\section{Introduction}
Spin glass physics has been widely studied over the last decades because
spin glasses are considered as the paradigm for investigating the 
`glass' state~\cite{SGreviews1,SGreviews2}.
In particular, the Ising Edwards-Anderson spin glass model defined by 
\be 
H_{\rm Ising} = - \sum_{\langle i,j \rangle} J_{ij} S_i S_j,
\label{iham}
\ee
has been very heavily studied~\cite{SGreviews1,SGreviews2,reviewsimu},
since it is the
simplest model with the necessary ingredients of randomness and frustration.
Here, the $S_i = \pm 1$ are Ising spins on a regular lattice
interacting through nearest neighbor interactions, $J_{ij}$, which are 
random variables drawn from a distribution 
of zero mean.
The poor theoretical understanding of issues such 
as the phase diagram of the Hamiltonian in Eq.~(\ref{iham}), 
the nature of its low temperature phase, or the 
extension of its mean-field solution to finite dimensions
shows that the problem is indeed challenging.
Also, due to the nature of the problem, experiments
only probe non-equilibrium dynamics of spin glasses 
at low temperature because 
the equilibration time of a macroscopic sample is infinite in this region. 
Experiments therefore pertain to the field of non-equilibrium 
statistical mechanics~\cite{review1,review1_1}.
The variety 
of dynamic phenomena observed in experiments (aging, 
rejuvenation, memory, etc)
can be viewed as additional theoretical 
challenges~\cite{review1,review1_1,reviewth,review3}.

In recent years, several theoretical approaches to the slow dynamics
of spin glasses described the physics in terms of a 
distribution of length scales whose time and temperature evolution depends
on the specific experimental protocol, leading to a
good qualitative understanding of the dynamics of spin 
glasses~\cite{reviewth,review3,fh,jp,encorejp,yosh,surf,yosh5_1,yosh5,ludo1,ludo1_1}. 
Early numerical studies~\cite{rieger} revealed the existence of 
a corresponding dynamic
correlation length separating small quasi-equilibrated and 
large non-equilibrated length scales.
The physical relevance of these length scales
was however critically discussed only more recently,
both in simulations~\cite{ludo1,ludo1_1,rieger,yosh2,BB} and in 
experiments~\cite{encorejp,dupuis,yosh4,yosh3}.

However, the connection between simulations and
experiments is unclear 
because the spins in an experimental spin glass 
have a vectorial character, so that a more natural 
Hamiltonian to consider is
\be
H = - \sum_{\langle i,j \rangle} J_{ij} {\bf S}_i \cdot {\bf S}_j,
\label{hham}
\ee
where the ${\bf S}_i$ are now three-component vectors of unit length. 
The Heisenberg Edwards-Anderson model in Eq.~(\ref{hham})
has been far less studied than its
Ising counterpart.
Experimentally, anisotropy induced by Dzyaloshinsky-Moriya
interactions allows the study of 
``Ising-like'' or ``Heisenberg-like'' samples, depending on its strength.
Recent experiments performed on Ising and Heisenberg
samples revealed that the distinction indeed  
matters~\cite{dupuis,yosh3,ian}. For instance, 
different samples behave quite differently even if similar
temperature protocols are used~\cite{dupuis,yosh3}.
This emphasizes the need for large scale studies of the 
non-equilibrium dynamics~\cite{ludo2} of the Hamiltonian in
Eq.~(\ref{hham}).

We have therefore
performed detailed  non-equilibrium
simulations of the Heisenberg Edwards-Anderson
spin glass model in three-dimensions. In this
paper we discuss results obtained following the simplest, 
yet widely studied, experimental protocol where the system is quenched 
at initial time from a high-temperature state to its 
spin glass phase.
The result of temperature shift and cycling experiments, and the influence
of finite  cooling rates on the dynamics are the object 
of a future paper~\cite{prep}. 

The paper is organized as follows. In 
Sec.~\ref{model}, we present the model and give technical
details. The dynamics following a quench is 
presented in Sec.~\ref{aging}. Scaling behavior 
of dynamic functions is discussed in Sec.~\ref{scaling}.
We give a summary of our results in the conclusion
of the paper in Sec.~\ref{conclusion}.

\section{Model and numerical details}
\label{model}

We numerically
study the model defined by the Hamiltonian in Eq.~(\ref{hham}), 
in which the Heisenberg spins lie on the sites 
of a three-dimensional cubic lattice with $N=L^3$ sites
periodic boundary
conditions. 
The random couplings are drawn from a Gaussian distribution 
of zero mean and standard deviation unity.
We use a heat-bath algorithm~\cite{peterold} in which 
the updated spin has the correct Boltzmann distribution for the instantaneous
local field. This method has the advantage that a change in the spin
orientation is always made. Times will be given 
in Monte Carlo sweeps,
where one Monte Carlo sweep 
represents $N=L^3$ spin updates.
We use a rather large simulation box of linear size $L=60$, 
and discuss below in more detail this choice for $L$.
We study several temperatures $T=0.16$, 0.15, 0.14,
0.12, 0.10, 0.08, 0.04 and 0.02.
Although all the quantities we shall study 
are self-averaging, we average over several 
realizations of the disorder, typically 15, to 
increase the statistics of our data. 

In this paper, we simulate a single type of thermal history, 
corresponding to ``simple aging experiments'', as opposed 
to the increasingly complex thermal protocols which have been 
proposed in recent years~\cite{review1,review1_1,review3}. 
In a simple aging experiment, the system is initially prepared 
in a high temperature state, $T \gg T_c$, and suddenly quenched
at initial time $t_w=0$ in the low temperature phase, $T < T_c$. 
The temperature is then kept constant throughout the 
experiments, while dynamical measurements are performed
at various ``waiting times'' $t_w$ after the quench.
Some ``two-time'' quantities correlating the system at
$t_w$ with that a time $t$ later are also determined.
We study the system for a total of $10^5$ sweeps (the largest value of
$t_w + t$) with
the following 20 values of $t_w$, which are in a roughly
logarithmic progression: 2, 3, 5, 9, 16, 27, 46, 80, 139, 240, 416, 720, 1245,
2154, 3728, 6449,
11159, 19307, 33405, 57797.

Contrary to its Ising version, even the location and nature
of the spin glass transition of the model in Eq.~(\ref{hham}) 
have been a matter of debate, although the situation has 
clarified somewhat recently. Early simulations 
reported the existence of a zero temperature critical
point~\cite{peterold,bana,mcmillan},
in plain contrast to experimental findings~\cite{SGreviews1,SGreviews2}.
Kawamura proposed to resolve 
this discrepancy by introducing
the spin-chirality decoupling scenario~\cite{kawa1,kawa,kawa2,kawa3}, 
based on Villain's ideas
that non-collinear ground states might exist in 
systems with vector spins~\cite{villain}. 
More recently, several papers~\cite{matsu,endoh,matsu2,peternew}
contradicted this scenario and argued, 
for both the Gaussian and $\pm J$ versions of the model (\ref{hham}),
that spins and chiralities in fact order at the same critical temperature,
$T_c>0$.
Very recent simulations~\cite{peternew} involving the most efficient 
tools used to study the Ising spin glass~\cite{ballesteros} conclude 
that the present model is characterized by a phase 
transition at $T_c \simeq 0.16$, where both
spins and chirality simultaneously freeze. 
This motivates the choice of $T = 0.16$
for the upper temperature in our simulations.

The present study has three main aims. 

(1) {\it Comparison with experiments.} As explained above, 
the model in Eq.~(\ref{hham}) is best suited to describe 
experimental samples where the spins also have a vectorial character. 
We want therefore to compare our numerical results to 
experimental findings. Since our work is the first large scale simulation
of the Heisenberg spin glass in the aging regime, we will 
intentionally display a wide range of numerical data covering the whole 
low temperature phase and various observables.

(2) {\it Comparison with the Ising spin glass.}
As mentioned above, experiments have revealed quantitative differences 
between Heisenberg and Ising samples. Therefore, we also 
want to compare our results with the many aging studies 
of the Ising EA model, Eq.~(\ref{iham}).

(3) {\it The nature of the transition in the Heisenberg model.}
Since some of the numerical works that support the spin-chirality
decoupling scenario
are performed in a dynamical context, we shall discuss
this issue in detail, investigating the gradual freezing with time
of both spin and chiral degrees of freedom. Following 
Ref.~\onlinecite{kawa1}, we define chirality as 
\be
\chi_i^\mu = {\bf S}_{i+{\bf e}_\mu} \cdot ({\bf S}_i \times 
{\bf S}_{i-{\bf e}_\mu} ),
\ee 
where $\mu \in \{ x, y, z \}$, and ${\bf e}_{\mu}$ is a unit vector 
in the direction $\mu$.

\section{Aging dynamics}
\label{aging}

In this section, we define and study the behavior of various 
dynamical quantities that are measured during the aging 
of the system. Their scaling properties are discussed in detail
in the following section.

\subsection{Energy density}

\begin{figure}
\begin{center}
\psfig{file=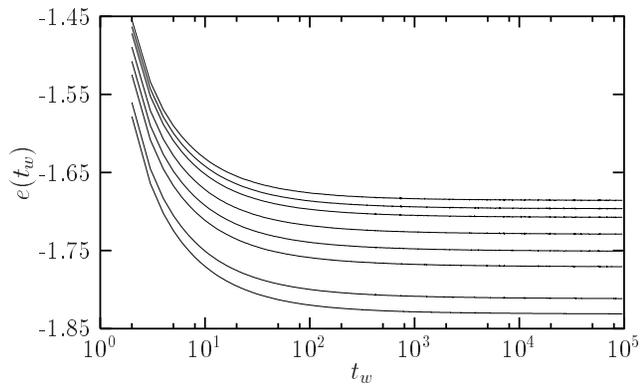,width=8.5cm}
\caption{\label{energy} 
The time dependence of the energy density for temperatures 
$T=0.16$, 0.15, 0.14, 0.12, 0.10, 0.08, 0.04 and 0.02 
(from top to bottom) reveals the aging of the system
in its low temperature phase, $T \le T_c \simeq 0.16$.}
\end{center}
\end{figure}

It is a central feature of glassy materials that they do not reach 
thermal equilibrium on experimental time scales when 
they are quenched to their ``glassy'' phase. 
The main consequence is that physical quantities keep evolving
with time as the system tries to reach equilibrium, which is known as
``aging'', a term invented by the polymer glass community~\cite{struik}. 

An obvious manifestation of this out-of-equilibrium 
dynamics is therefore the time dependence
of physical observables. In our case, it is easy to follow 
the evolution of the energy density, 
\begin{equation}
e(t_w) = \left\langle \frac{1}{N} H \right\rangle.
\end{equation}
That the system ages at all temperatures studied here
is indeed clear from Fig.~\ref{energy} where at each 
temperature $T\in [0.02, 0.16]$ the time dependence of 
$e(t_w)$ is evident. 

\subsection{Two-time autocorrelation functions}

\label{twotime}
The evolution with time of physical observables 
implies that the dynamics of the system is not time translationally invariant.
Early studies on polymeric glasses showed that two-time quantities
reveal the aging behavior of the system much more strikingly~\cite{struik}, 
so that two-time correlation or response functions are 
widely studied in aging glassy materials.

\begin{figure}
\begin{center}
\psfig{file=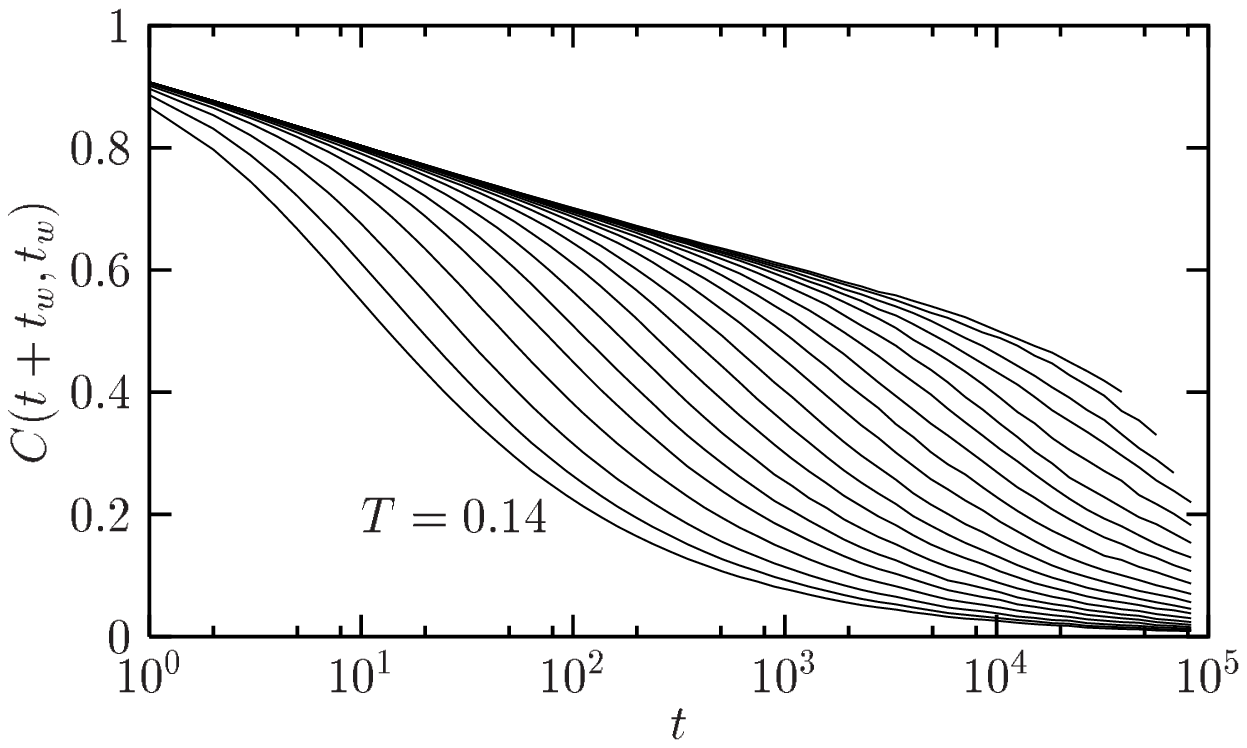,width=8.5cm} 
\psfig{file=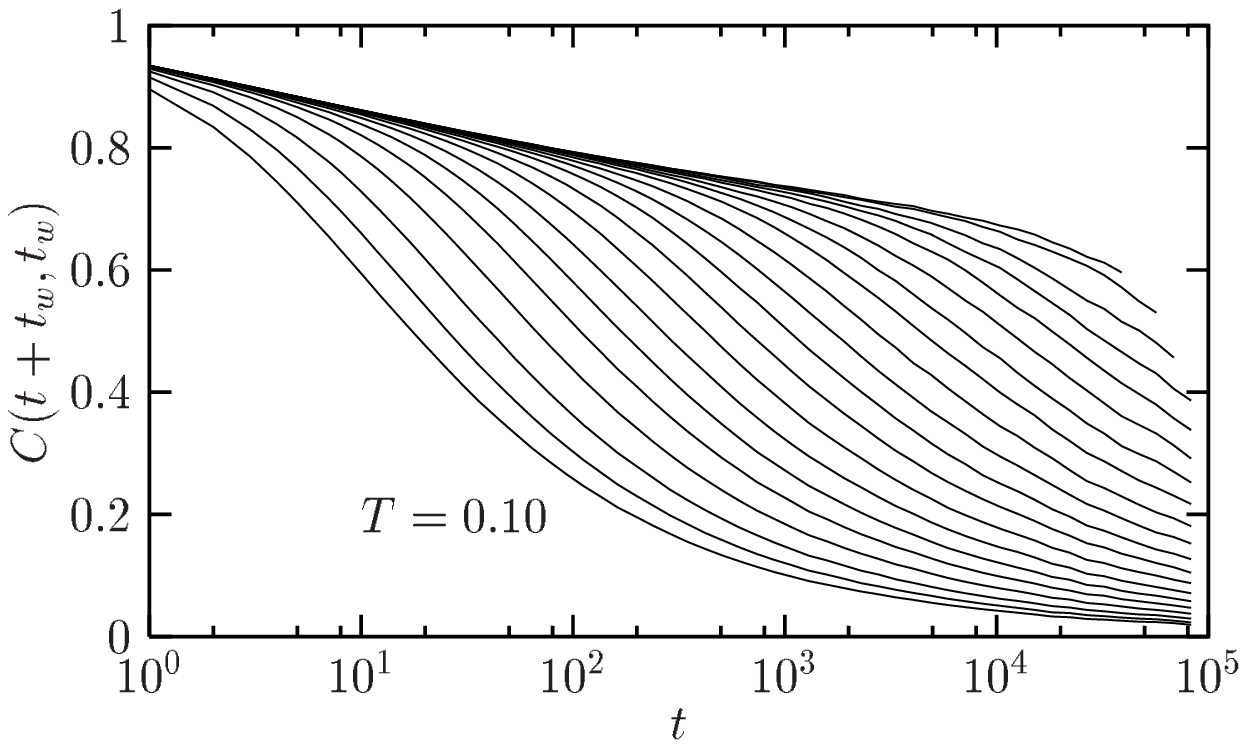,width=8.5cm} 
\psfig{file=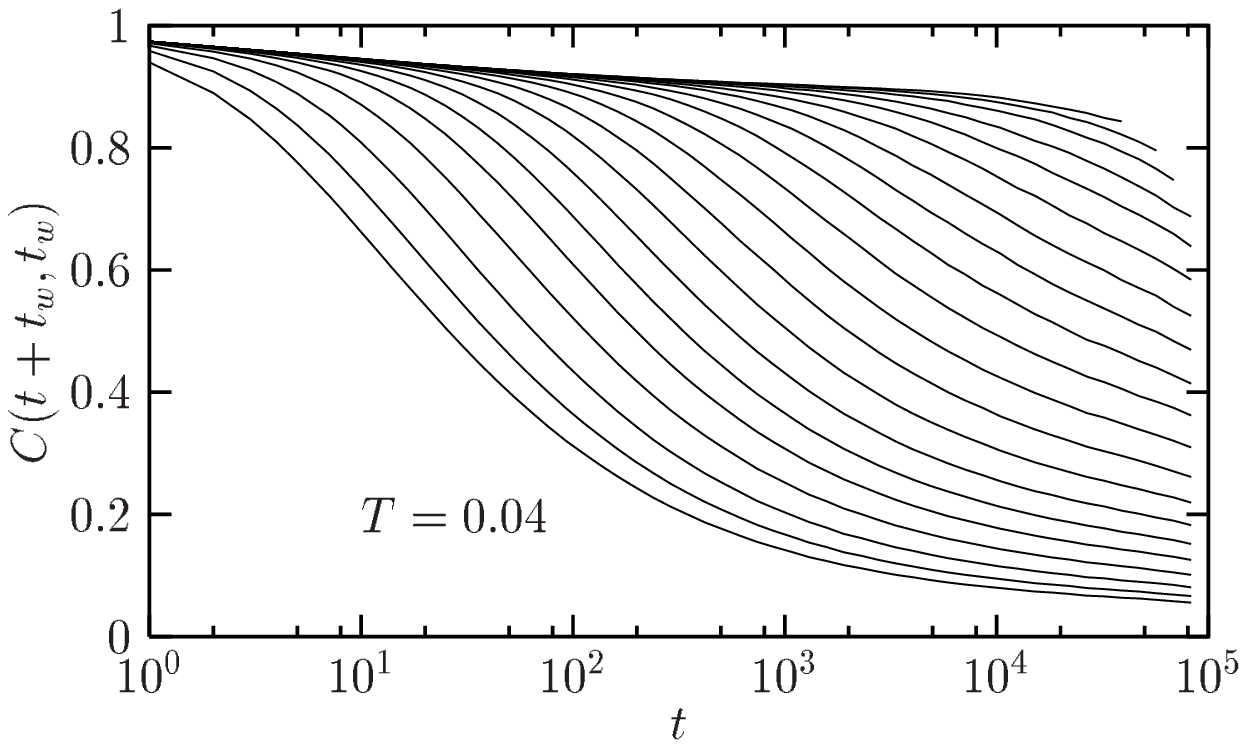,width=8.5cm} 
\caption{\label{auto} Autocorrelation function
of the spins, Eq.~(\ref{autospin}),
for $L=60$, as a function
of the time difference $t$ for various waiting times $t_w$
logarithmically spaced in the interval $t_w \in [2, 57797]$ (from left
to right). 
The temperature is $T=0.14$, 0.10 and 0.04 (from top to 
bottom).}
\end{center}
\end{figure}

The simplest two-time quantity that has been studied 
numerically in spin glasses is the autocorrelation function 
of the spins defined by 
\begin{equation}
C(t+t_w,t_w) = {1\over N} \sum_i
\langle {\bf S}_i (t+t_w) \cdot {\bf S}_i (t_w) \rangle.
\label{autospin}
\end{equation}
This function is represented as a function of the time difference $t$
for various waiting times $t_w$ and various temperatures in 
Fig.~\ref{auto}.

The first and main observation from Fig.~\ref{auto} is that 
this two-time quantity is not a function of the time difference 
only, as it would be in an equilibrium system, 
$C(t+t_w,t_w) \ne C(t)$. This is just another way of saying that
the system is aging, but much more information can be extracted
from this correlator.

In more detail, the shape of the curves shown in Fig.~\ref{auto}
is similar to what is commonly observed in many materials. 
For small time differences, $t \ll t_w$, the curves for various
$t_w$ superpose, implying that the dynamics is time translationally invariant
in this time regime, reminiscent of some sort of ``local'' 
or ``quasi'' equilibrium. This will be clarified below.
Moreover, we find that for all $T \le 0.10$, the
curves in this time regime are consistent 
with the existence of a ``plateau''. 
At $T=0.10$, the plateau starts to be visible
at time differences $t \sim 10^3 - 10^4$ only, but its existence 
becomes clearer 
at lower temperatures.
In mathematical terms, this 
means that the Edwards-Anderson parameter defined as
\be
q_{\rm EA}  = \lim_{t \to \infty} \lim_{t_w \to \infty} \lim_{L \to\infty}
C(t+t_w,t_w), 
\label{ea}
\ee
is finite and positive, $q_{\rm EA} > 0$.
It is important that the system is large enough that it can be considered
effectively infinite for the values of $t$ and $t_w$ used.
If the system is not
large enough, the overall direction of the spins will wander randomly during
the simulation, since
there is no energy cost to a global rotation, with the result that
$\lim_{t \to \infty} \lim_{t_w \to \infty}C(t+t_w,t_w) = 0$ at
fixed $L$. 
It is clear 
from Fig.~\ref{auto} that $q_{\rm EA}$ is a decreasing
function of temperature, with $q_{\rm EA}(T \to 0) = 1$. 
At higher temperatures, we do not observe a plateau, presumably 
because our time window is too small, as can be guessed from
comparing the curves at $T=0.14$ and $T=0.10$ in Fig.~\ref{auto}.

Although a plateau is expected, because equilibrium measurements indicate spin
glass order below $T_c$, 
we recall that no such plateau can be unambiguously 
observed in the Ising spin glass in three dimensions, where
the short time regime is well-described by a pure power law~\cite{rieger,BB}. 
A non-zero $q_{\rm EA}$ indicates the existence
of spin glass phase where spins are frozen in random directions, and
the observation of a plateau in the spin autocorrelation function in
Fig.~\ref{auto}
is the first evidence of a standard spin glass phase
in the model in Eq.~(\ref{hham})
using non-equilibrium techniques.  For the Ising spin glass,
analogous evidence for a transition is currently missing.
Experimentally, plateaus are also hardly visible in
two-time correlation or response functions, but this is
probably due to the narrow experimental time window.
The existence of a plateau can be, however, experimentally revealed
through a scaling analysis of two-time
quantities~\cite{review1,review1_1}.

Turning to the large time regime, $t \gg t_w$, we 
observe that curves at various waiting times do not superpose at 
all in this regime at any temperature, fully
revealing the aging nature of the dynamics. 
As in many glassy materials, it is clear that the time decay
of $C(t+t_w,t_w)$ becomes slower when the waiting time
increases. 
The physical interpretation
is simple: since the relaxation time 
of the sample is infinite, the only relevant 
time scale is the age of the sample $t_w$ which
imposes an age-dependent relaxation time: the older the sample, 
the slower its relaxation becomes. 
We shall discuss in the next
section the scaling behavior of these curves.

\begin{figure}
\begin{center}
\psfig{file=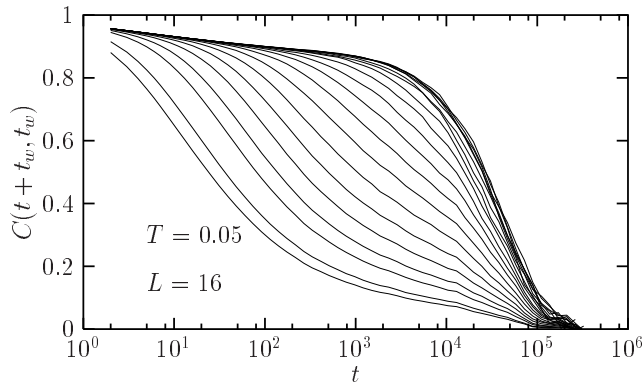,width=8.5cm}
\caption{\label{autoL16} Autocorrelation function
of the spins for $T=0.05$ and $L=16$ show that
the use of too small a system size yields stationary data
for $t_w > 10^4$, not observed in the larger system used
in Fig.~\ref{auto}.}
\end{center}
\end{figure}

\begin{figure}
\begin{center}
\psfig{file=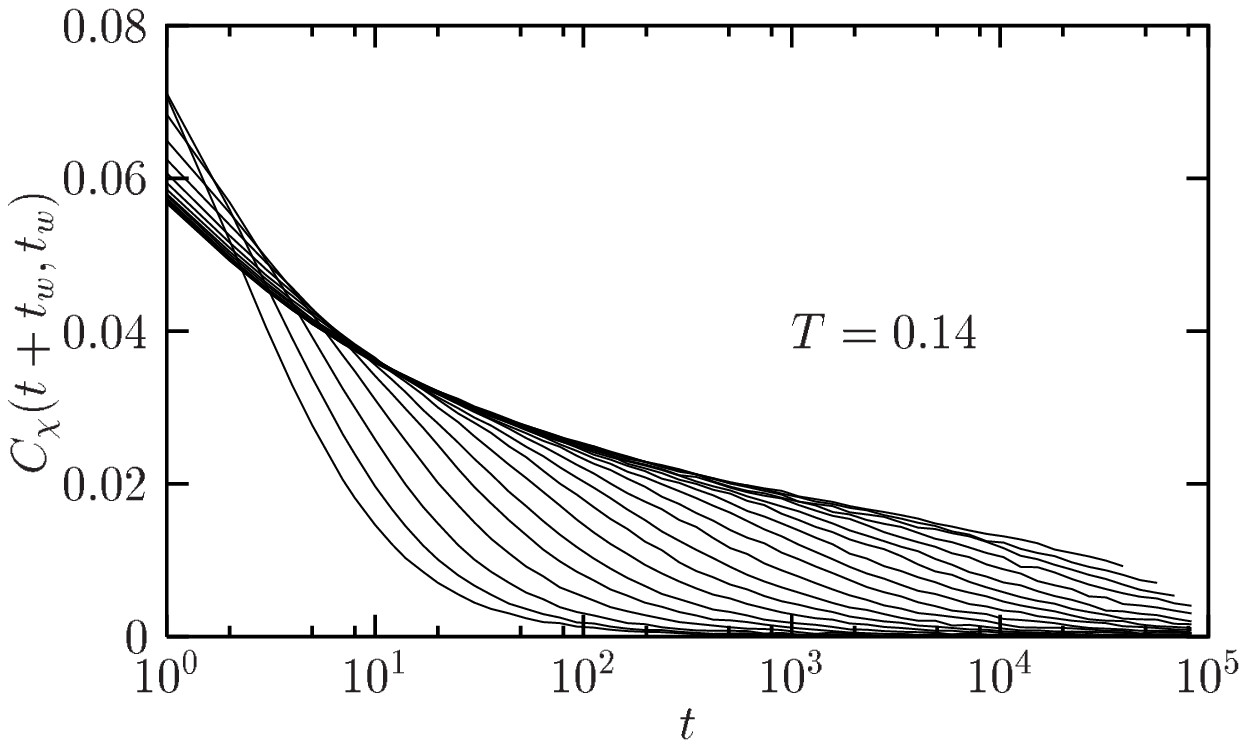,width=8.5cm} 
\psfig{file=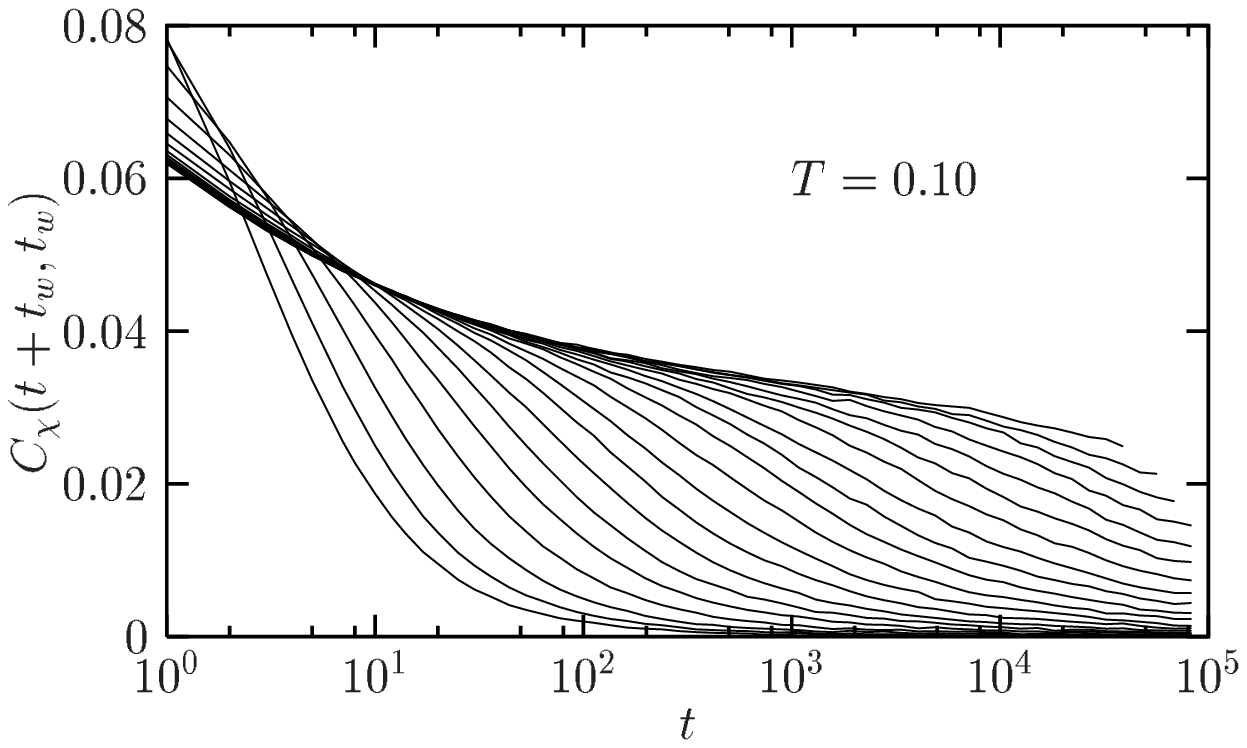,width=8.5cm} 
\psfig{file=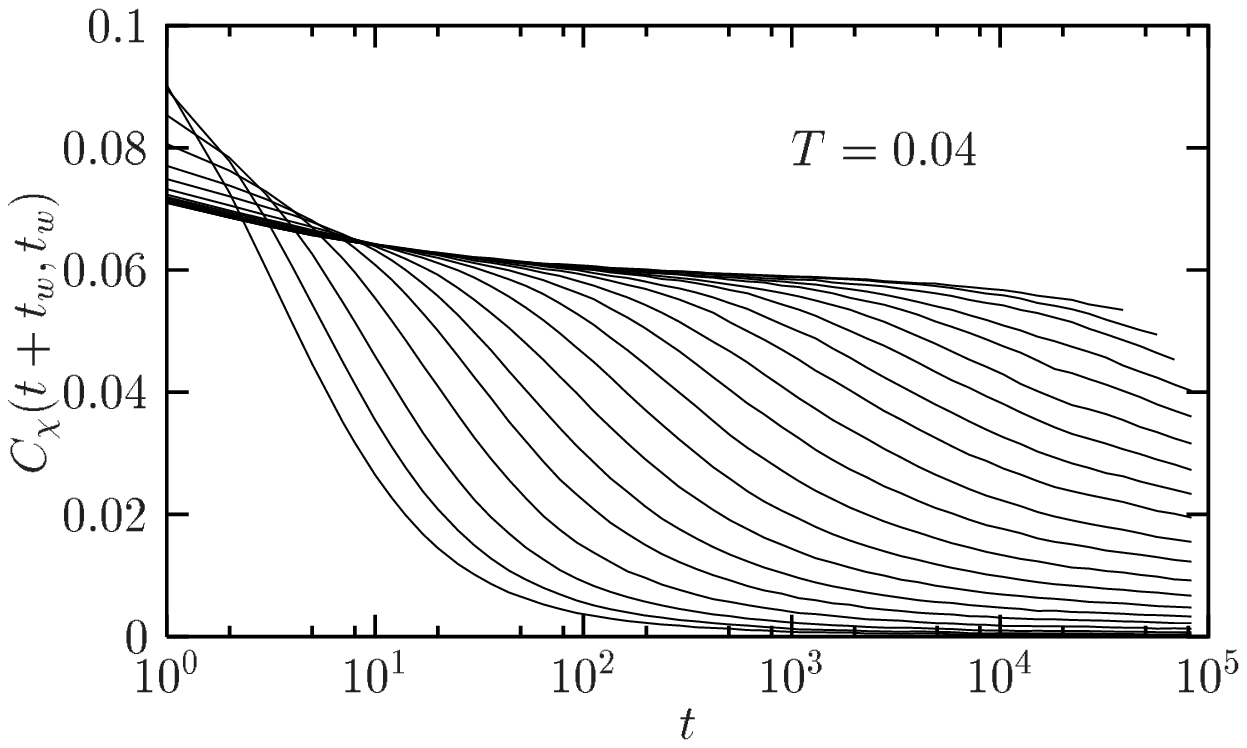,width=8.5cm}
\caption{\label{autochir}Autocorrelation function
of the chirality, Eq.~(\ref{autochireq}), as a function
of the time difference $t$ for various waiting times $t_w$
logarithmically spaced in the interval $t_w \in [2, 57797]$ (from left to
right).
The temperature is $T=0.14$, 0.10 and 0.04 (from top to 
bottom).}
\end{center}
\end{figure}

The long time behavior found here is qualitatively 
very different from the one reported by Kawamura~\cite{kawa} who argues 
that the spin autocorrelation function becomes stationary 
at large times, even at temperatures as low as $T=0.05$. 
This fact was later corrected by Matsubara {\it et al.}~\cite{matsu}
who noted that a global rotation or ``drift'' of the system could affect 
the dynamics, and produced curves similar to ours 
by ``subtracting'' by hand a global rotation of the spins. 
This is more simply interpreted as the system being too small for the order of
limits in Eq.~(\ref{ea}) to apply.
Our results for an $L=16$ system, plotted
in Fig.~\ref{autoL16}, are consistent
with those of Kawamura~\cite{kawa}, and show stationary behavior at long
times.
By contrast, for $L=60$, 
the spin autocorrelation function shown in Fig.~\ref{auto}, does 
not reach stationarity in the same time window even
for temperatures as high as $T=0.16$ and without subtracting
a global rotation of the spins.
This shows that one of Kawamura's numerical
arguments in favor of a spin-chirality decoupling~\cite{kawa} stems from 
data on too small a size.
Similar finite-size effects 
are most probably also at work in Refs.~\onlinecite{kawa2,kawa3}. 

We now present data for the autocorrelation function of the chirality, 
\be
C_\chi(t+t_w,t_w) = \frac{1}{N} \sum_i
\langle \chi_i^\mu (t+t_w) \chi_i^\mu (t_w) \rangle.
\label{autochireq}
\ee
Since the system is isotropic, $C_\chi$ does not depend on $\mu$ and 
we have also averaged the data over the three directions of space.
Our results are shown in Fig.~\ref{autochir} for the same 
parameters as for the spins. The main conclusion from Fig.~\ref{autochir}
is that chiralities 
have essentially the same behavior as the spins. We observe
a stationary decay at small $t$, followed by a slower, waiting time dependent
decay at large times. As for the spins, the appearance of a plateau 
is clear within our time window for $T \le 0.10$, indicating 
the existence of a non-zero Edwards-Anderson parameter
for chirality, 
\be
q_{\chi {\rm EA}} = \lim_{t \to \infty} \lim_{t_w \to \infty} 
\lim_{L\to\infty}
C_\chi(t+t_w,t_w),
\ee
which also grows when $T$ decreases. This is expected since we have 
found above that spins freeze, which implies 
that chiralities freeze as well.

\subsection{Time is length}

\begin{figure}
\begin{center}
\psfig{file=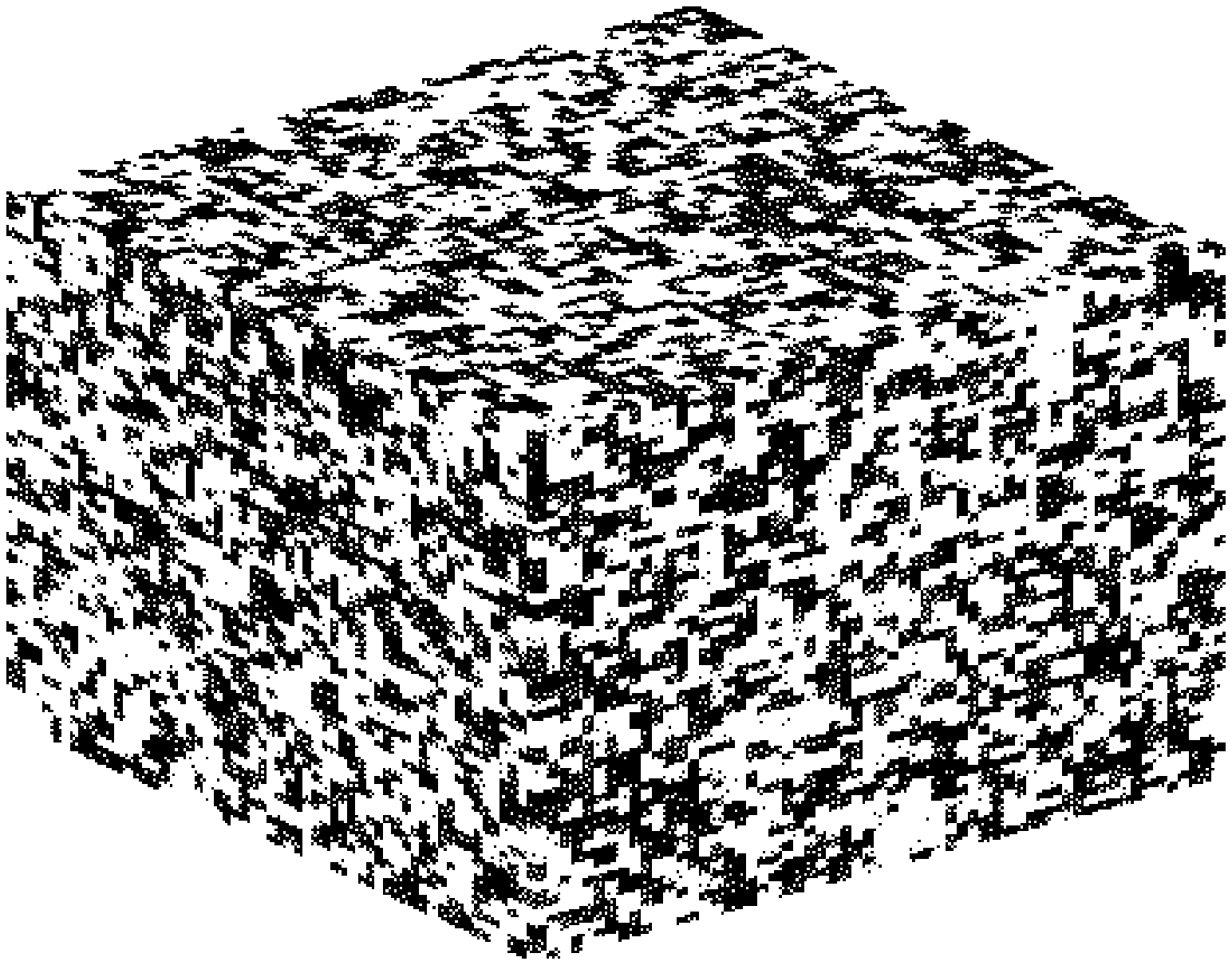,width=5.25cm,height=5.25cm} 
\psfig{file=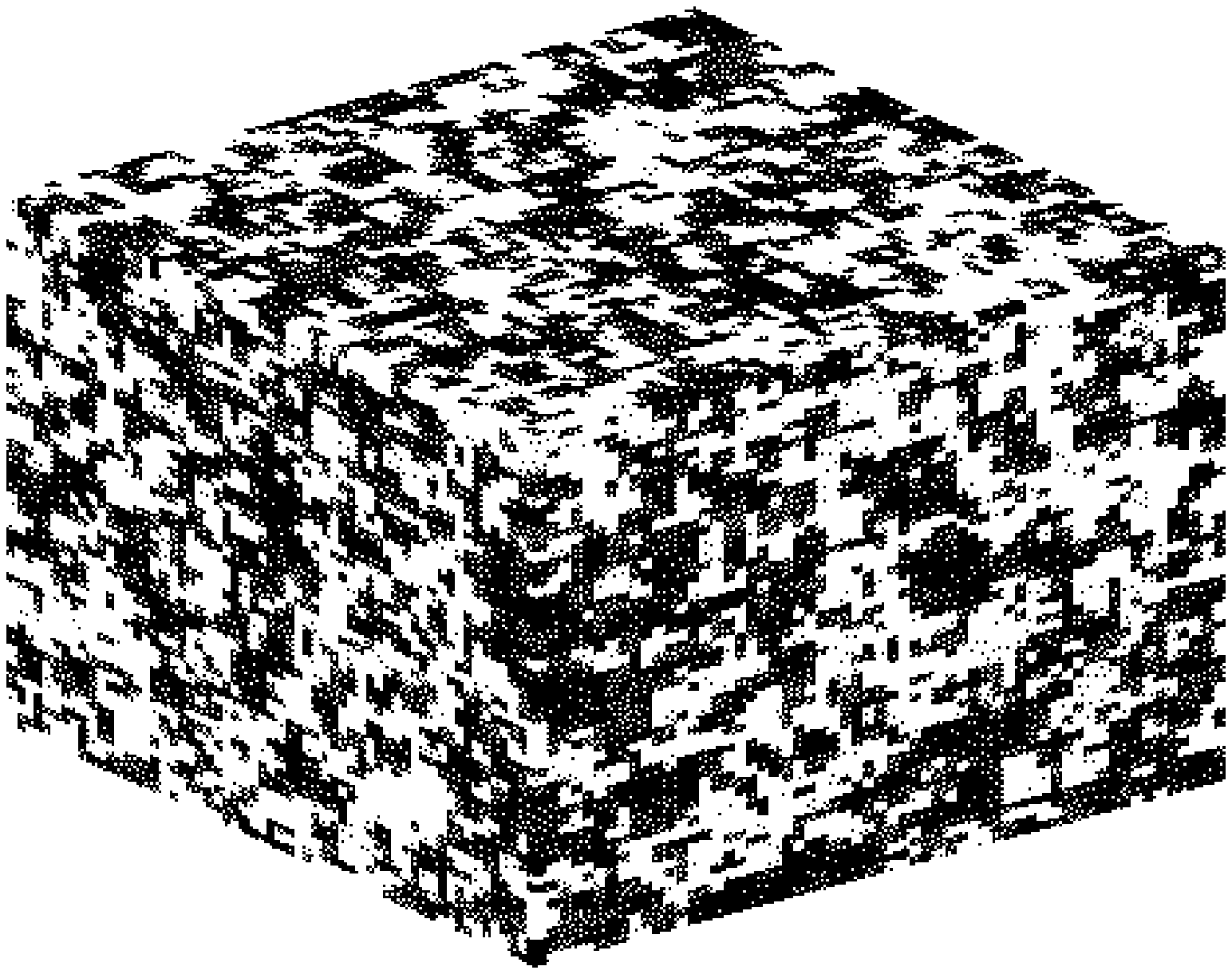,width=5.25cm,height=5.25cm} 
\psfig{file=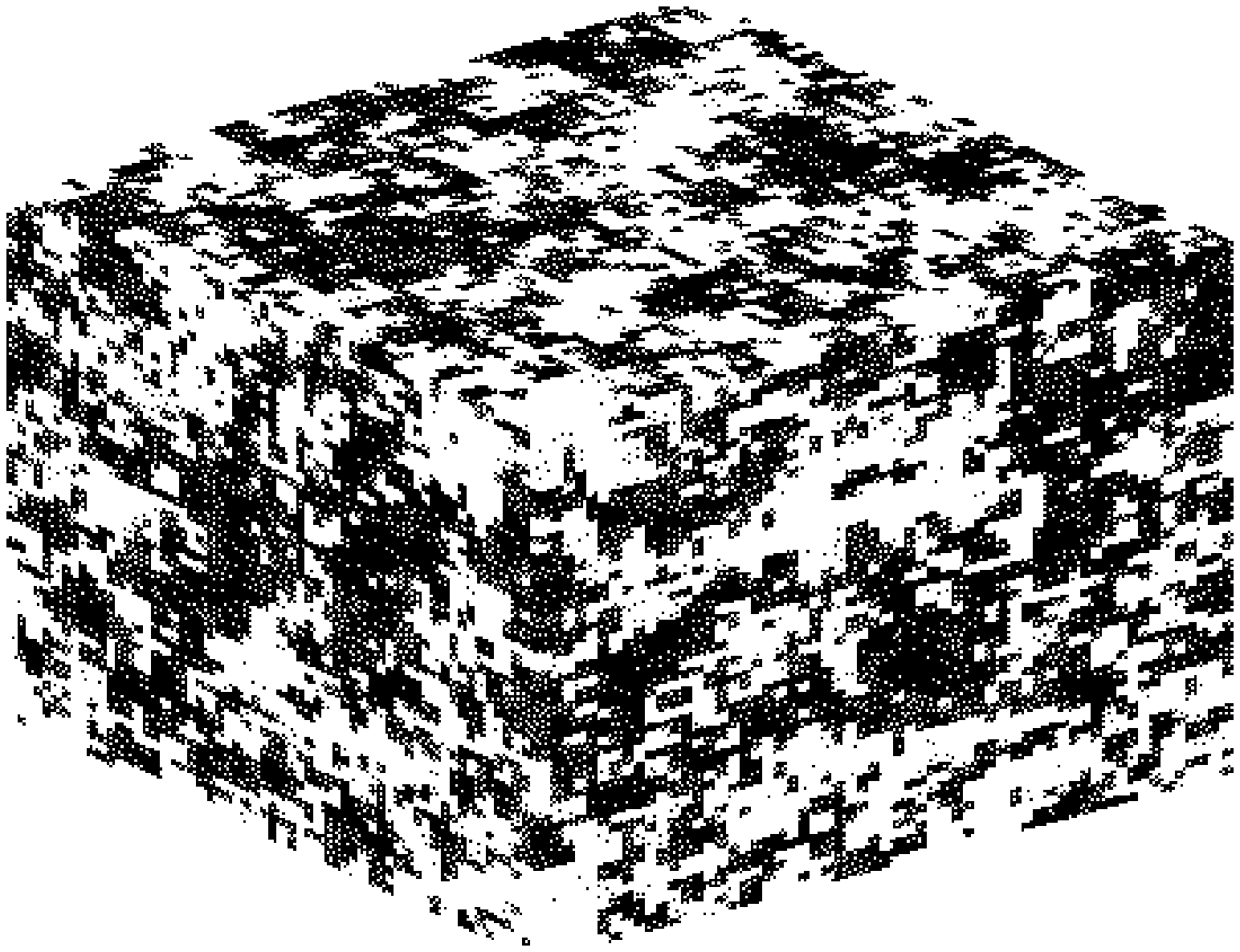,width=5.25cm,height=5.25cm} 
\caption{\label{pic} The relative orientation of the spins
in two copies of the system, Eq.~(\ref{cos}), is encoded on a grey scale
in a $60 \times 60 \times 60$ simulation box at three different
waiting times $t_w=2$, 27 and 57797 (from top to bottom) at 
temperature $T=0.04$. The growth of a local random ordering 
of the spins is evident.}
\end{center}
\end{figure}

\begin{figure}
\begin{center}
\psfig{file=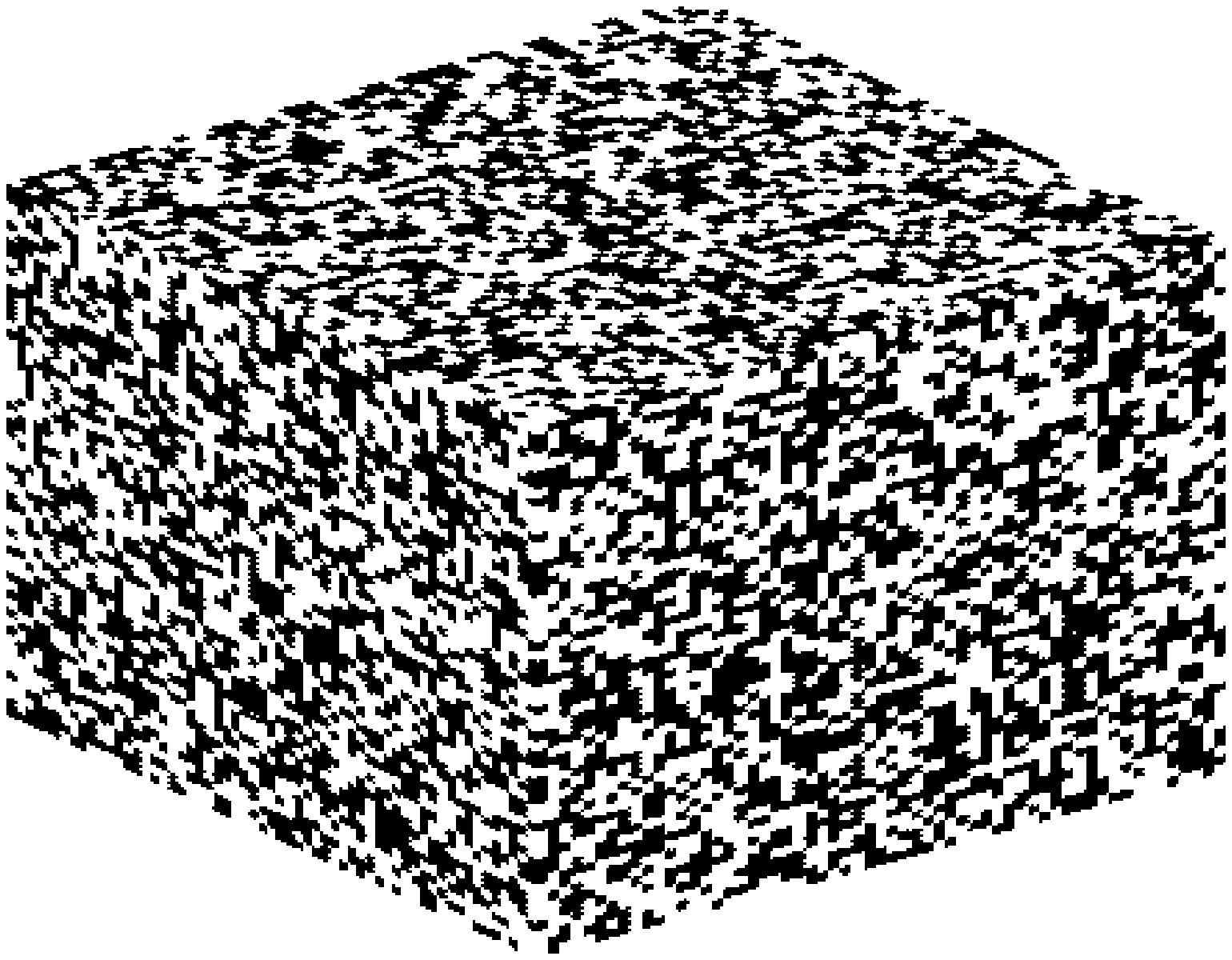,width=5.25cm,height=5.25cm} 
\psfig{file=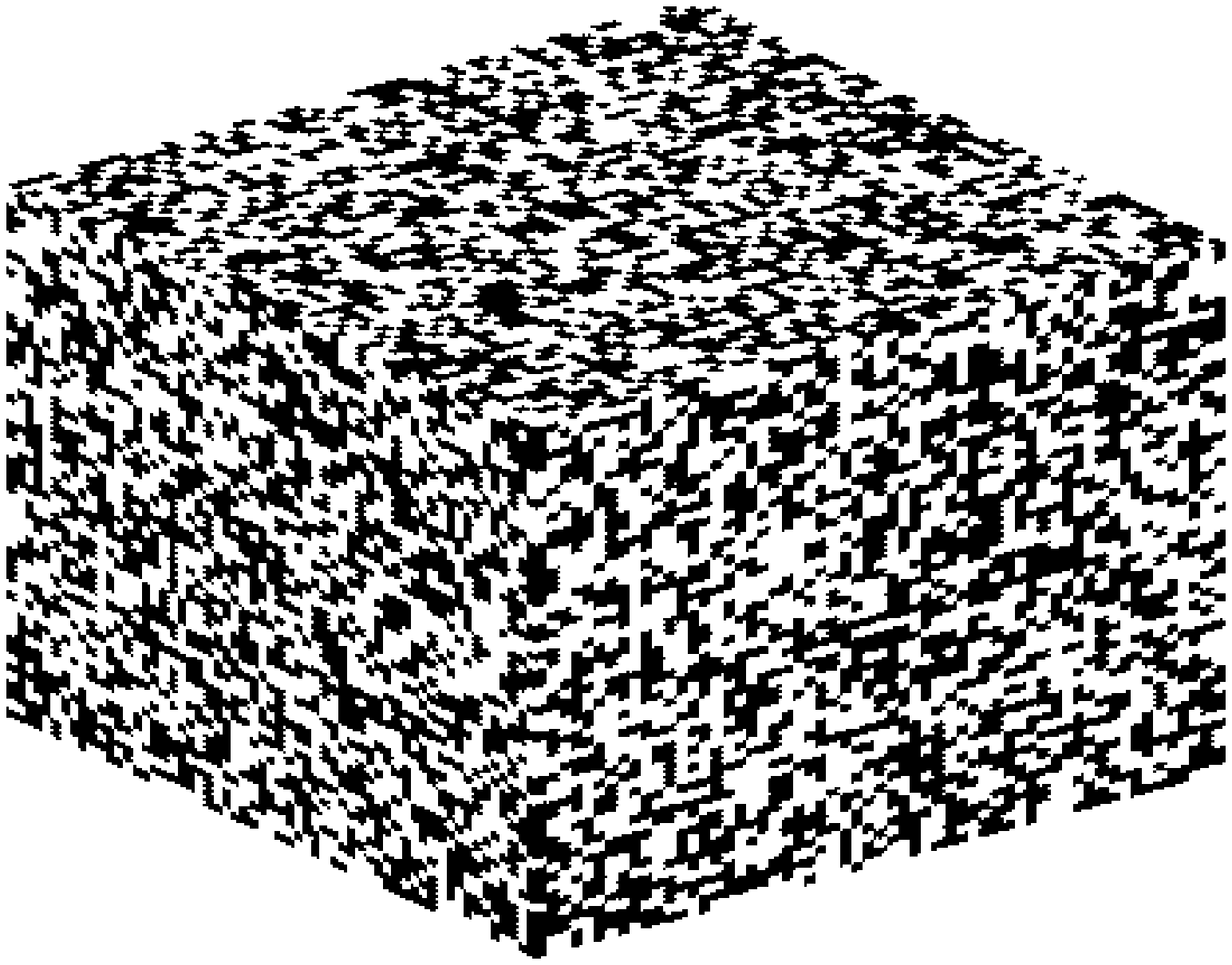,width=5.25cm,height=5.25cm} 
\psfig{file=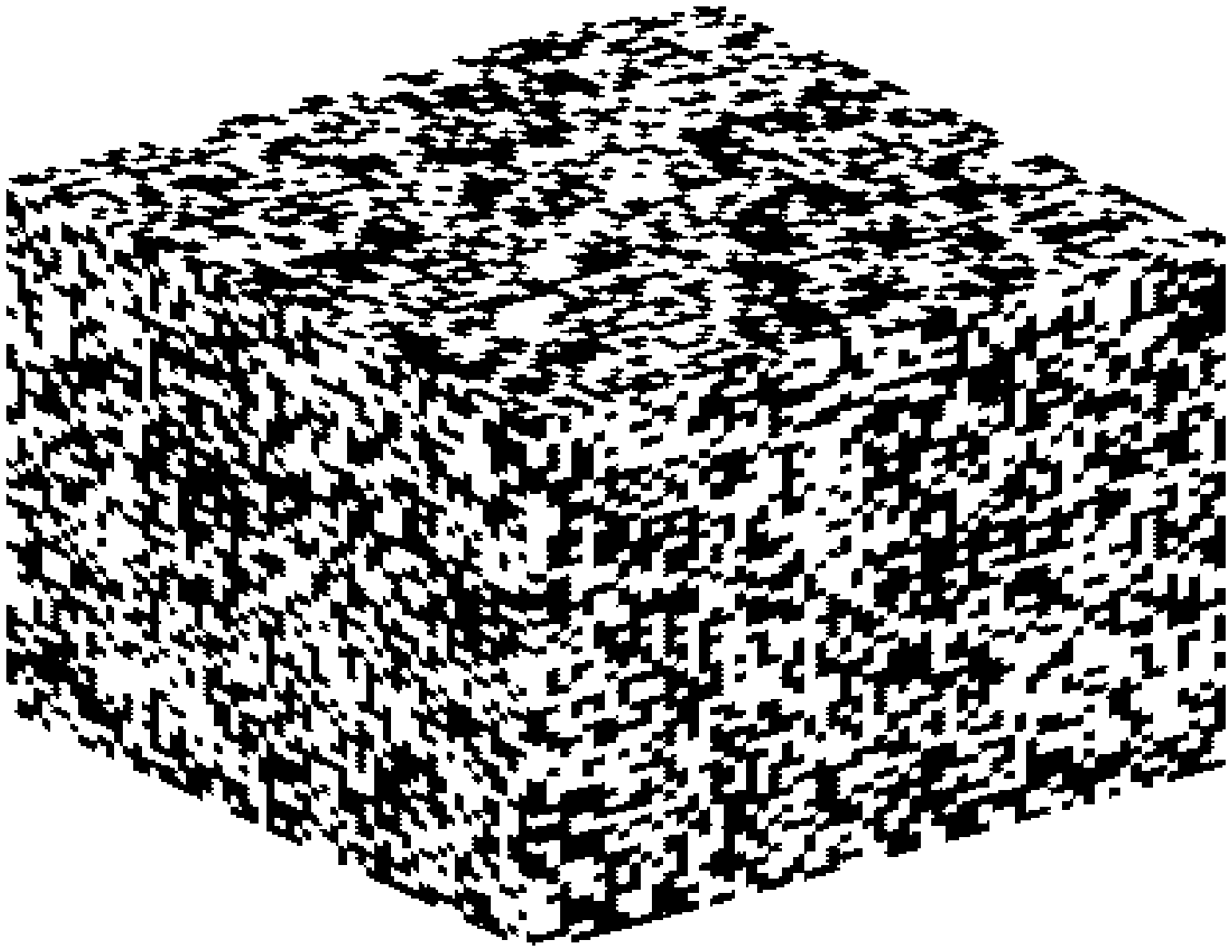,width=5.25cm,height=5.25cm} 
\caption{\label{picchir} The sign of the chiral local 
overlaps in two copies of the system, Eq.~(\ref{coschir}), is encoded on a 
black and white scale in a $60 \times 60 \times 60$ simulation box at 
three different waiting times $t_w=2$, 27 and 57797 (from top to bottom).
The temperature is $T=0.04$, and we have chosen $\mu = x$.
The actual configurations are the same
as in Fig.~\ref{pic}, but the chiral
ordering, corresponding to the spin ordering observed there,
is hardly visible.}
\end{center}
\end{figure}

The key problem is to understand the subtle slow 
changes that the system undergoes: what does ``old'' or ``young'' 
really mean for the sample? To answer this question, we turn
to a spatial description of the aging dynamics.

The decomposition of the decay
of autocorrelation functions into
a fast stationary process and a slow
non-stationary one directly suggests the existence of some 
sort of local equilibrium
within the sample: a spin appears locally
equilibrated (short-time dynamics) although the sample as a whole
is still far from equilibrium and evolves towards
equilibrium (long-time dynamics). 

It is possible to illustrate this last statement, as was 
done in the Ising case~\cite{rieger}. Because of the disorder, 
the spin orientations in an equilibrium
configuration are random, so that it is impossible to
detect any domain growth by simply looking at the spin
directions. However, two copies of the system,
$a$ and $b$, evolving independently but with
the same realization of the disorder, will reach correlated
equilibrium configurations, so that the orientation
of the spins in one copy can be compared 
with those in the second copy.

It is therefore useful to define the local relative orientation 
of the spins as
\begin{equation}
\cos \theta_i (t_w) = {\bf S}_i^a (t_w) \cdot {\bf S}_i^b (t_w).
\label{cos}
\end{equation}
In Fig.~\ref{pic} we present snapshots where this quantity 
is encoded on a grey scale. Comparing three successive
times, it becomes clear that aging is nothing but the
growth with time of a local random ordering of the spins imposed
by the disorder of the Hamiltonian~\cite{fh,huse}.
Notice that the ``domains''
observed in Fig.~\ref{pic} have highly irregular boundaries, which
will influence the behavior of the spatial
correlators discussed below.

Next we focus
on chiral degrees of freedom, and similarly define 
a chiral local overlap as 
\be
q_{\chi i}^\mu (t_w) = \chi_i^{\mu, a} (t_w) \chi_i^{\mu, b} (t_w).
\label{coschir}
\ee
In Fig.~\ref{picchir}, we present snapshots where this quantity 
is encoded on a black and white scale, i.e. we represent 
the quantity ${\rm sgn} (q_{\chi i}^\mu)$ for $\mu = x$. Different 
space directions would give similar plots.
Although a chiral ordering must follow the spin ordering observed 
in Fig.~\ref{pic}, this is hardly visible by the eye and the 
system appears much more disordered in this chiral representation. 
We interpret this as stemming from the fact that  
spins are actually not ``very'' correlated
within the dynamic correlation length (see the next subsection), 
and so the chiralities, which involve three spins, are even less correlated.

\subsection{Four-point correlation functions}

We now go beyond qualitative pictures of black and white domains and
measure the dynamic correlation length 
associated with 
the mean domain size observed in Figs.~\ref{pic} and \ref{picchir}.

First, we generalize the two-site, two-replica correlation
function (which is therefore a ``four-point'' object)
studied in the Ising case~\cite{rieger} 
to the case of Heisenberg spins as
\begin{equation}
C_4(r,t_w) = {1 \over N}
\sum_i\langle {\bf S}_i^a (t_w)  \cdot {\bf S}_{i+r}^a 
(t_w) \,\, {\bf S}_i^b (t_w) \cdot {\bf S}_{i+r}^b 
(t_w) \rangle.
\label{c4eq}
\end{equation}
This function measures correlations of the relative 
orientation of two spins separated by a distance $r$ at time $t_w$, 
just as the structure factor does in a pure ferromagnet. 
Note that
$C_4(r,t_w)$ is invariant under global rotation of the spins in either copy,
and so is independent of the wandering of the overall spin orientation which
can affect the two-time autocorrelation functions discussed in
Sec.~\ref{twotime}. However, there will be a change of behavior in
$C_4(r,t_w)$ when $t_w$ is sufficiently large that the dynamic
correlation length becomes comparable to the system size $L$, since the system
then equilibrates.

\begin{figure}
\begin{center}
\psfig{file=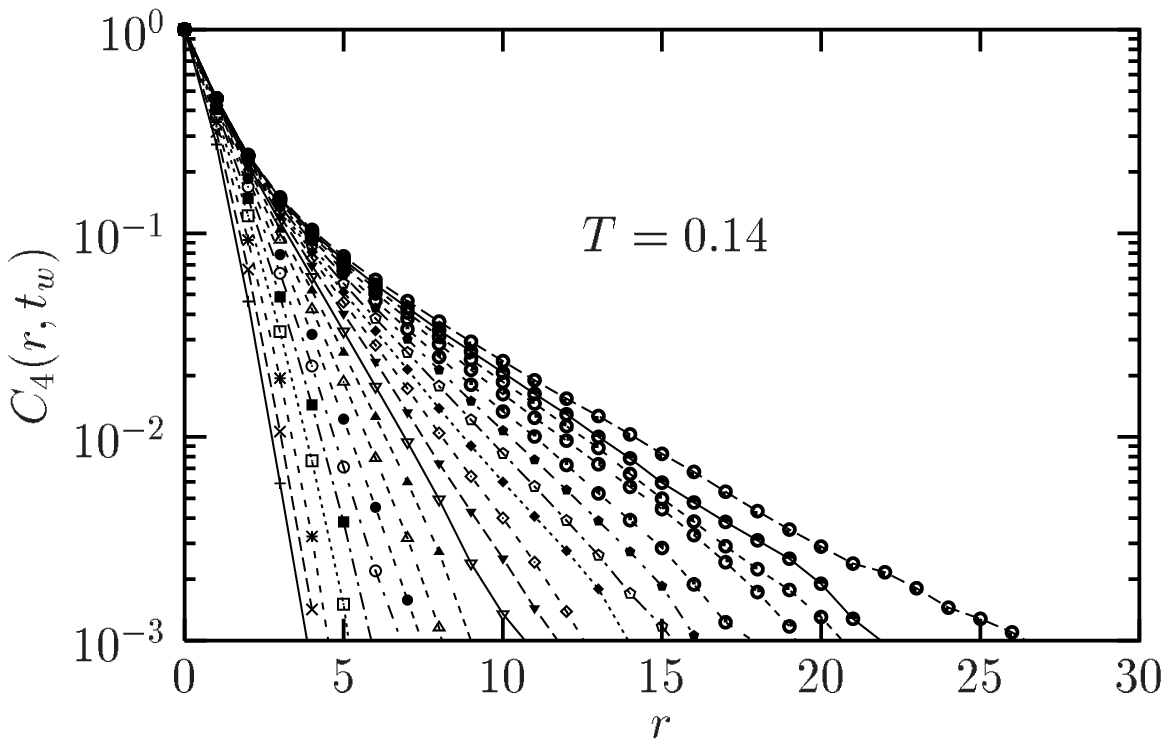,width=8.5cm} 
\psfig{file=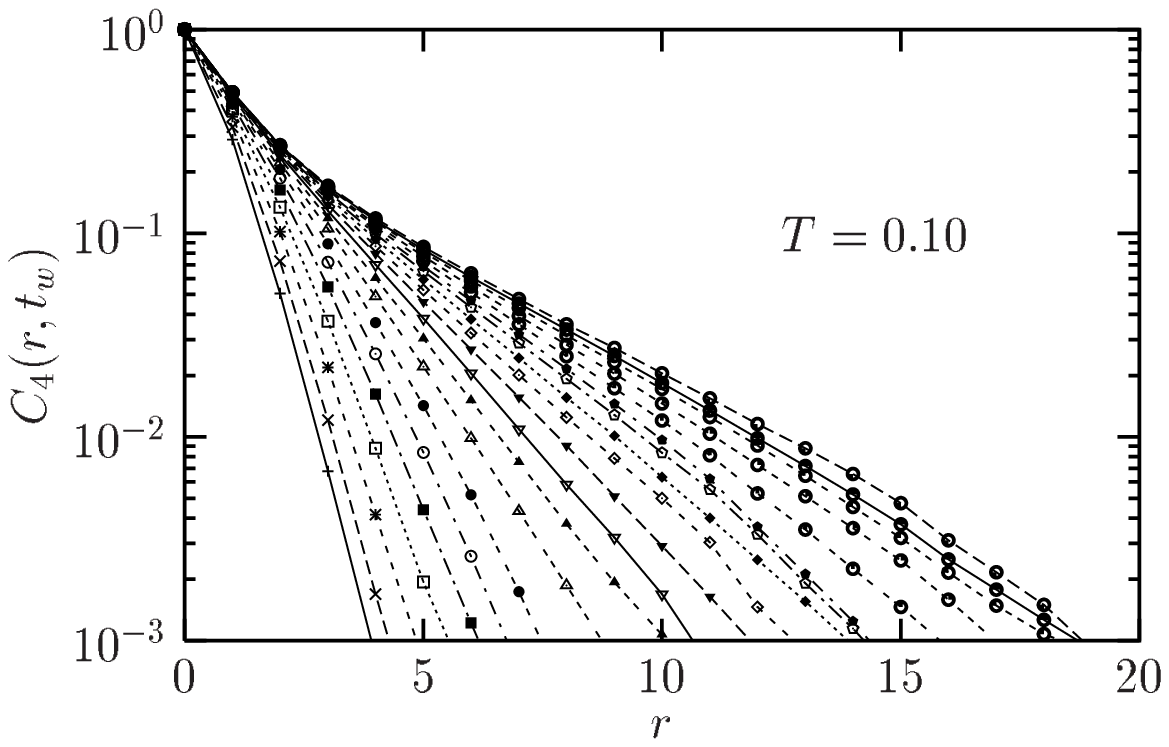,width=8.5cm} 
\psfig{file=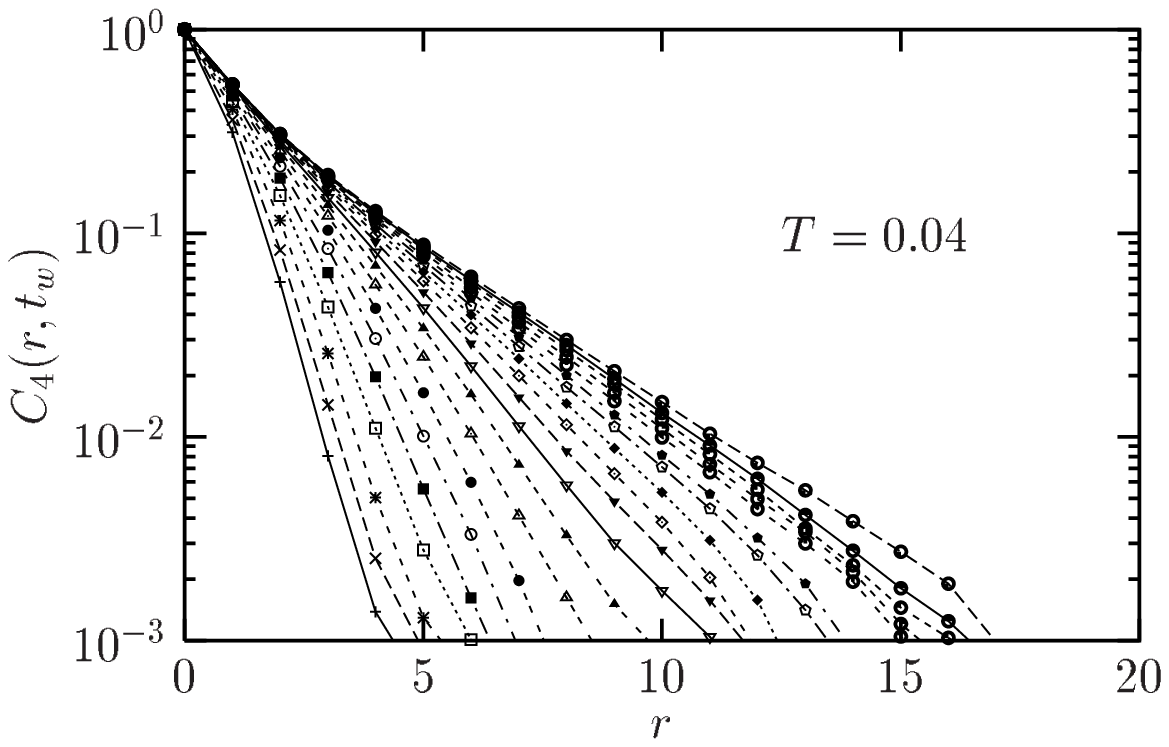,width=8.5cm} 
\caption{\label{c4} The two-site two-replica correlator for the spins defined
in Eq.~(\ref{c4eq}) as a function of $r$ for various waiting
times logarithmically spaced in the interval 
$t_w \in [2, 57797]$ (from left to right). 
The temperature is $T=0.14$, 0.10 and 0.04 (from top to bottom).
Note that the range of the $r$ axis changes with temperature
as a result of a slower growth of the dynamic correlation length 
at lower temperature. 
A non-exponential decay is also evident from these curves.}
\end{center}
\end{figure}

We present the space dependence of $C_4(r,t_w)$ for various $t_w$
and three different temperatures in Fig.~\ref{c4}.
These data obviously confirm the visual impression of the snapshots
in Fig.~\ref{pic}. At a given temperature, the decay 
of $C_4(r,t_w)$ with $r$ becomes slower at larger $t_w$, indicating
the growth with time of a dynamic correlation length, $\ell(T,t_w)$, 
sometimes also referred to as a ``coherence length''.  
Physically, this means that an ``older'' system exhibits
slower dynamics because of a larger dynamic correlation, 
very much as in standard coarsening phenomena~\cite{fh}. 

A second piece
information we get from Fig.~\ref{c4} is that the growth
of $\ell(T,t_w)$ is strongly dependent on temperature, since 
much larger length scales can be equilibrated at $T=0.14$ than 
at $T=0.04$. This is expected in a disordered system where
thermal activation is likely to play a role, and 
we shall quantify this statement in the next section.
We also note that much smaller length scales are reached 
in the same time window in the Ising spin glass, both in three
and four dimensions, where plots similar to Fig.~\ref{c4} typically 
stop~\cite{rieger,BB} at $r=5$--$10$, 
instead of $r=20$--$30$ used here.

\begin{figure}
\begin{center}
\psfig{file=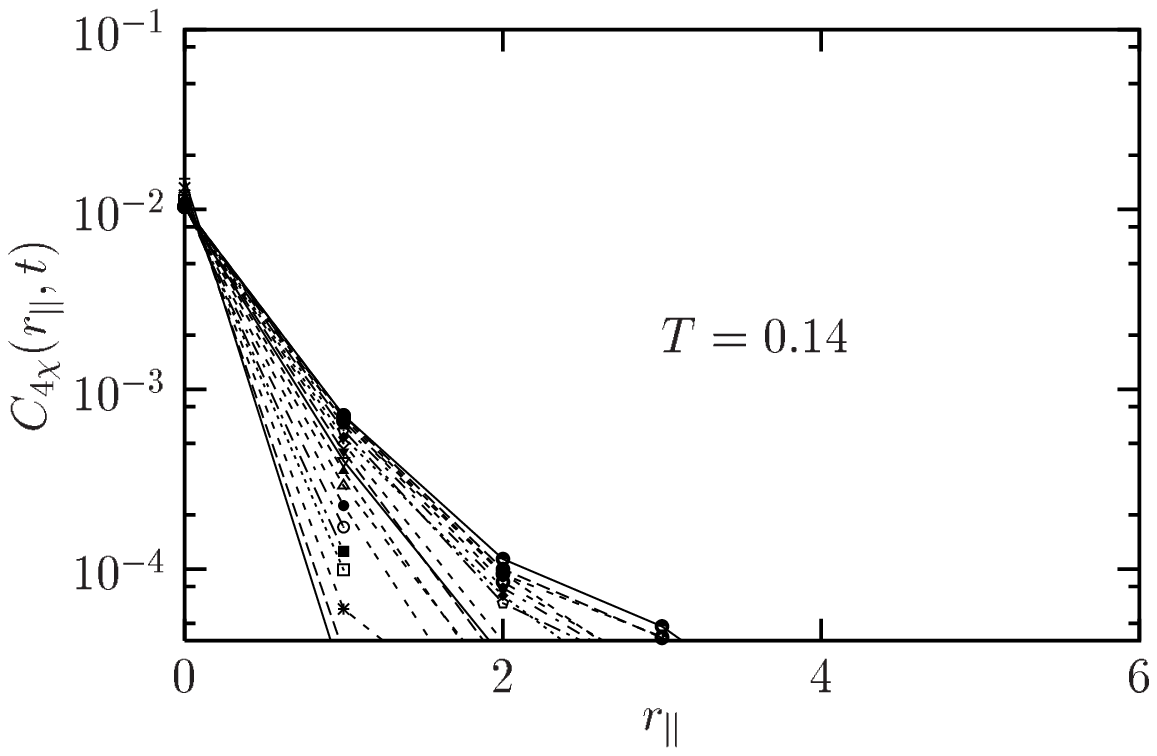,width=8.cm} 
\psfig{file=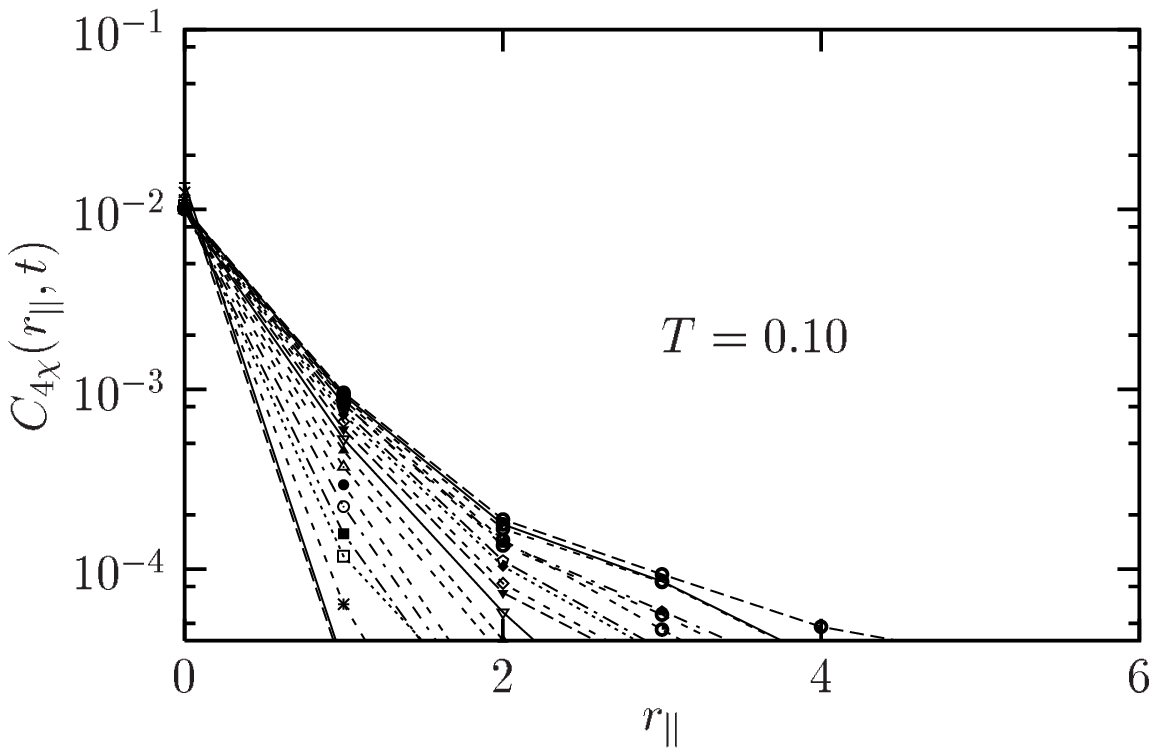,width=8.cm} 
\psfig{file=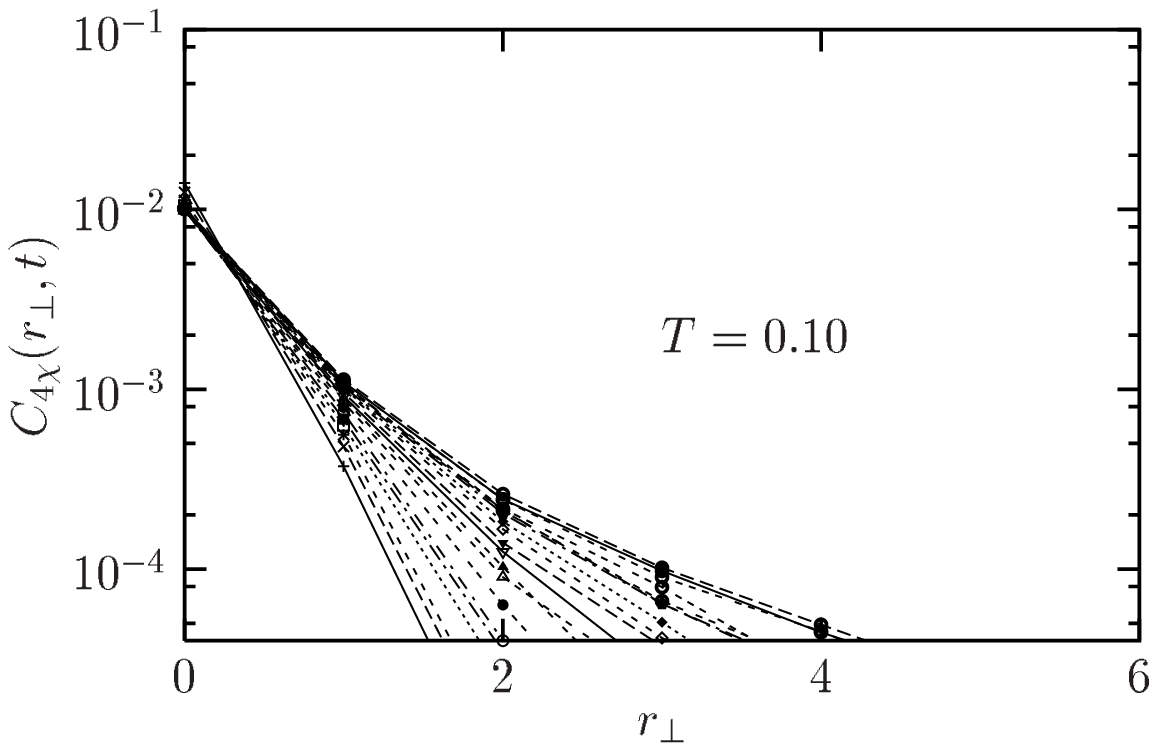,width=8.cm}
\psfig{file=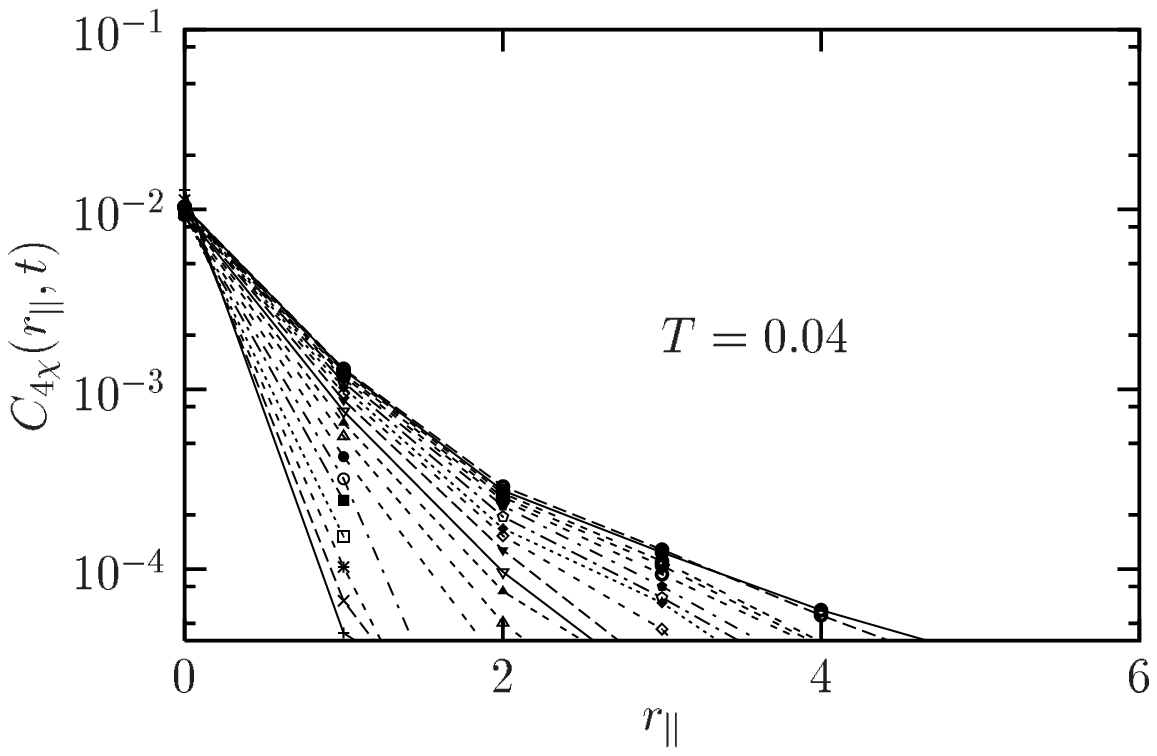,width=8.cm}
\caption{\label{c4chir}
The two-site two-replica correlator for the chiralities
(\ref{c4chireq}) as a function of $r$ for various waiting
times logarithmically spaced in the interval $t_w \in [2, 57797]$. 
The temperature is $T=0.14$, 0.10 and 0.04 (from top to bottom).
For $T=0.10$ we show both $C_{4 \chi}(r_\parallel,t_w)$ and 
$C_{4 \chi}(r_\perp,t_w)$ which exhibit similar behavior.
Note that the range of the $r$ axis is much
smaller than that for the spin correlator in Fig.~\ref{c4}.}
\end{center}
\end{figure}

A third piece of
information is that the spatial decay of $C_4(r,t_w)$
is clearly not exponential, since the latter would correspond to 
straight lines in the lin-log representation adopted in Fig.~\ref{c4}.
Moreover, a closer inspection of the data shows that, as for the time
decay of the autocorrelation functions, they can be decomposed 
in a short-distance decay, $r < \ell(T,t_w)$ where the curves
at various $t_w$ merge, and a long-distance one, $r > \ell(T,t_w)$,
which becomes slower at larger $t_w$. This confirms the intuition that
spins are indeed in local equilibrium on short length scales, but that
the system as a whole is not equilibrated. 
We note that even the local equilibrium part of the decay seems
to be non-exponential, which might be connected to 
the strongly irregular nature of the domains
observed in Fig.~\ref{pic}. 

Finally note that the correlator in Eq.~(\ref{c4eq})
is symmetric about $L/2 = 30$ due to periodic boundary conditions.
A look at Fig.~\ref{c4} justifies our use of a system size $L=60$ since 
even for the largest waiting time and the largest 
temperature studied here, we are in the regime where $\ell(T,t_w) < L/2$, 
so that our results are not affected by finite size effects. 

We now turn to the chiral degrees of freedom and define
corresponding two-site, two-replica spatial correlations 
of the chirality as 
\begin{equation}
C_{4 \chi}(r,t_w) = {1 \over N} \sum_i
\langle \chi_i^{\mu, a} (t_w)   \chi_{i+r}^{\mu, a} (t_w)
\chi_i^{\mu, b} (t_w)  \chi_{i+r}^{\mu, b}(t_w) \rangle.
\label{c4chireq}
\end{equation}
In the following we distinguish between two correlators 
$C_{4 \chi}(r_\parallel,t_w)$ and $C_{4 \chi}(r_\perp,t_w)$
if $r$ is taken in a direction parallel or perpendicular 
to $\mu$, respectively.

Our results for the correlators (\ref{c4chireq}) 
are presented in Fig.~\ref{c4chir}. As for the spins, 
the spatial decay becomes less fast at larger $t_w$, indicating
a gradual random ordering of the chiralities. Although
this ordering was not visible on the snapshots presented
in Fig.~\ref{picchir}, appropriate correlators 
not surprisingly perform better than the eye. 
Chiral ordering is anyway expected 
since ordering of the spins implies the one of chiralities. 
In agreement with the visual observations, however, 
we find that spatial correlations of chiralities are much weaker 
than for the spins and correlators are numerically
indistinguishable from noise beyond $r \sim 4$. 
As a result, we did not attempt to perform 
a detailed scaling analysis of spatial correlations
of the chirality.
Again, we interpret this as being due to the fact that 
chiralities are less correlated because they involve three 
spins on a length scale $r=2$.

\begin{figure}
\begin{center}
\psfig{file=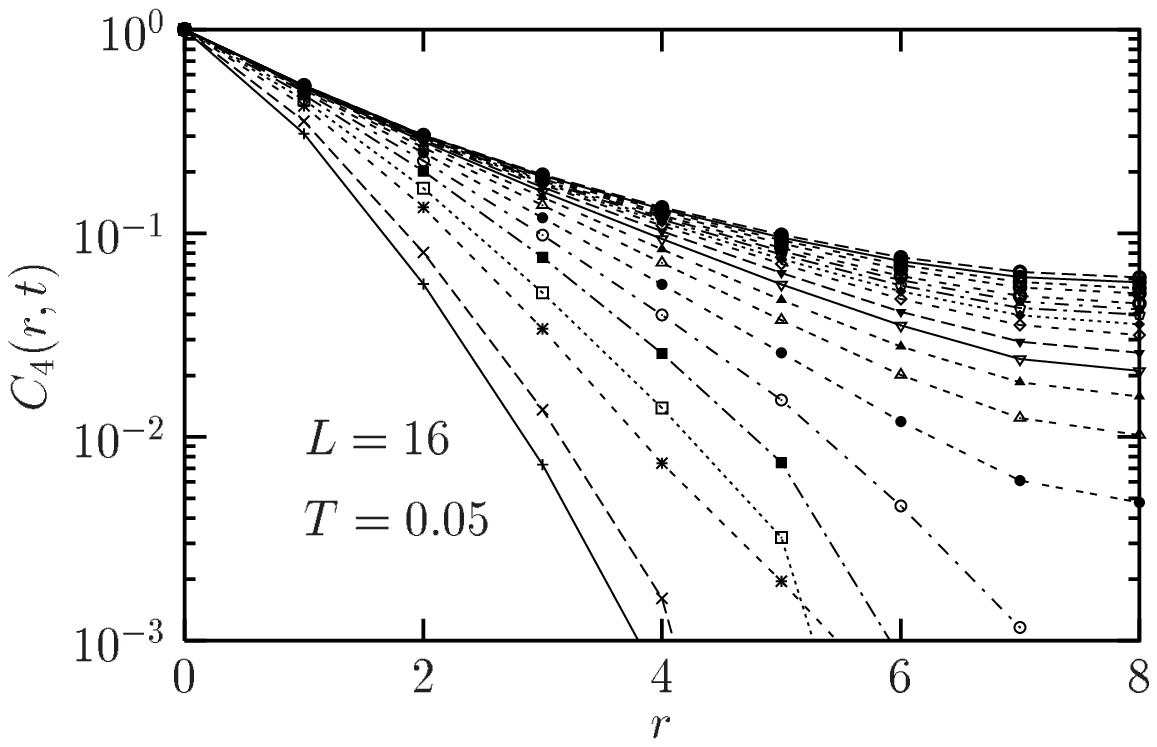,width=8.5cm} 
\psfig{file=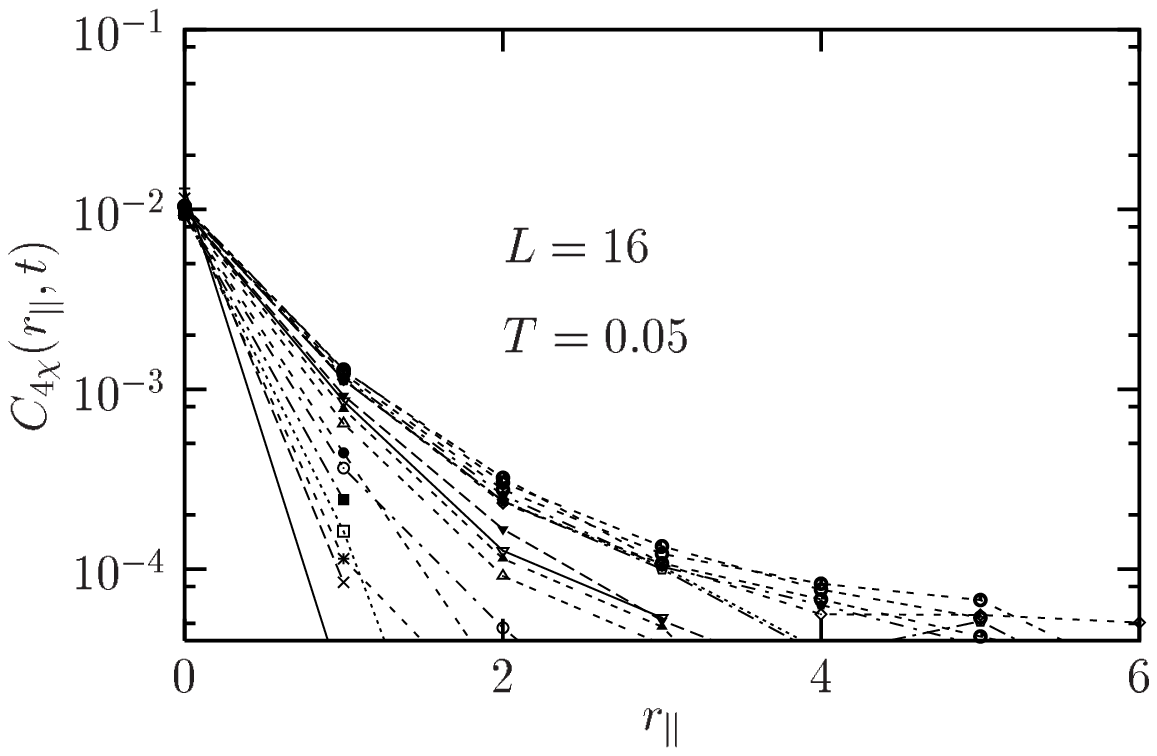,width=8.5cm} 
\caption{\label{c4L16} Two-site, two-replica functions for spins (top)
and chiralities (bottom) at $T=0.05$ and $L=16$. While 
the bottom figure is very similar to Fig.~\ref{c4chir}, the top
figure is very different (note in particular the different $r$-range used 
in both cases).}
\end{center}
\end{figure}

Results for the spin and chiral two-site, two-replica correlators for $L=16$,
the size studied by Kawamura~\cite{kawa}, are shown in Fig.~\ref{c4L16}.
By comparing the top of Fig.~\ref{c4L16} with Fig.~\ref{c4} which is for
$L=60$, we see that the
behavior of the spin function is very strongly size dependent. Physically this
is because the smaller system size comes to, or approaches, equilibrium on the
time scale of the simulation, whereas the larger size does not. However, the
data for the chiral function for $L=16$ in the bottom part of Fig.~\ref{c4L16}
is not very different from that for $L=60$ shown in Fig.~\ref{c4chir}.
This observation explains why Kawamura~\cite{kawa} found
that  chiral autocorrelation does not stop aging while 
spin autocorrelation does.

Although we have not been able to precisely estimate the dynamic
correlation length associated with chiral order, the fact that 
we cannot measure correlations beyond $r \sim 4$, while we can 
estimate spin correlations up to $r \sim 25$ shows that 
spin correlations are much stronger in the whole temperature
range that we have investigated, $T \leq 0.16$. This implies
that we find no temperature regime below $T=0.16$ where
chiral order manifests itself unaccompanied by 
simultaneous spin ordering as a spin-chirality decoupling scenario
naturally predicts.
Hence,
although the present non-equilibrium approach says nothing about 
equilibrium behavior in the thermodynamic limit, 
an important conclusion of this whole section is that,
when a proper system size is used, dynamical studies 
of the Heisenberg spin glass are more simply interpreted in terms 
of a simultaneous phase transition at $T_c \simeq 0.16$  
for both spin and chiral degrees of freedom.

\section{Scaling of dynamic functions}
\label{scaling}

The study of several space-time correlators 
of the previous section leads to the conclusion that
for $T \le 0.16$, spins of the Heisenberg spin glass 
gradually freeze with time in random orientations dictated by the quenched 
disorder, naturally followed by chiral degrees of freedom.
This behavior is qualitatively 
similar to that of the Ising spin glass. 
In this section, we study the scaling behavior of dynamic
functions defined for the spin degrees of freedom, 
to get a quantitative description of its aging dynamics, 
and compare the results with numerical studies
of the Ising spin glass and with experiments.

\subsection{Spatial correlations}

\begin{figure}
\begin{center}
\psfig{file=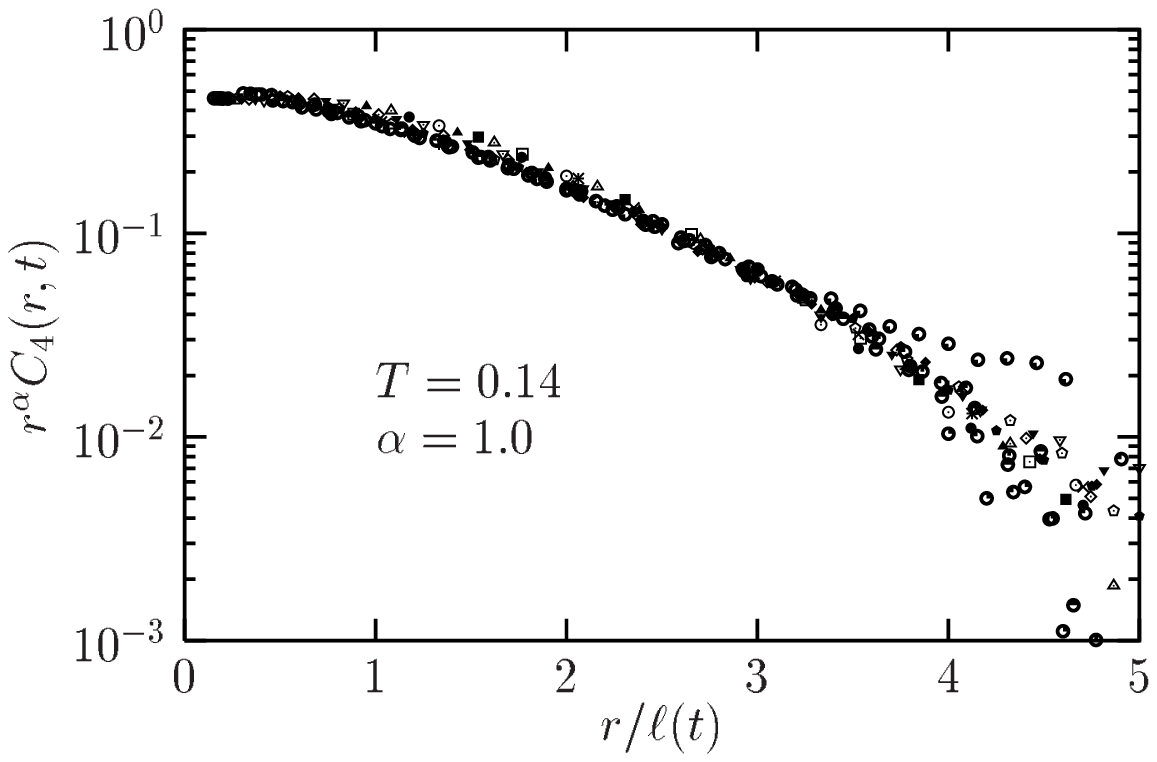,width=8.5cm} 
\psfig{file=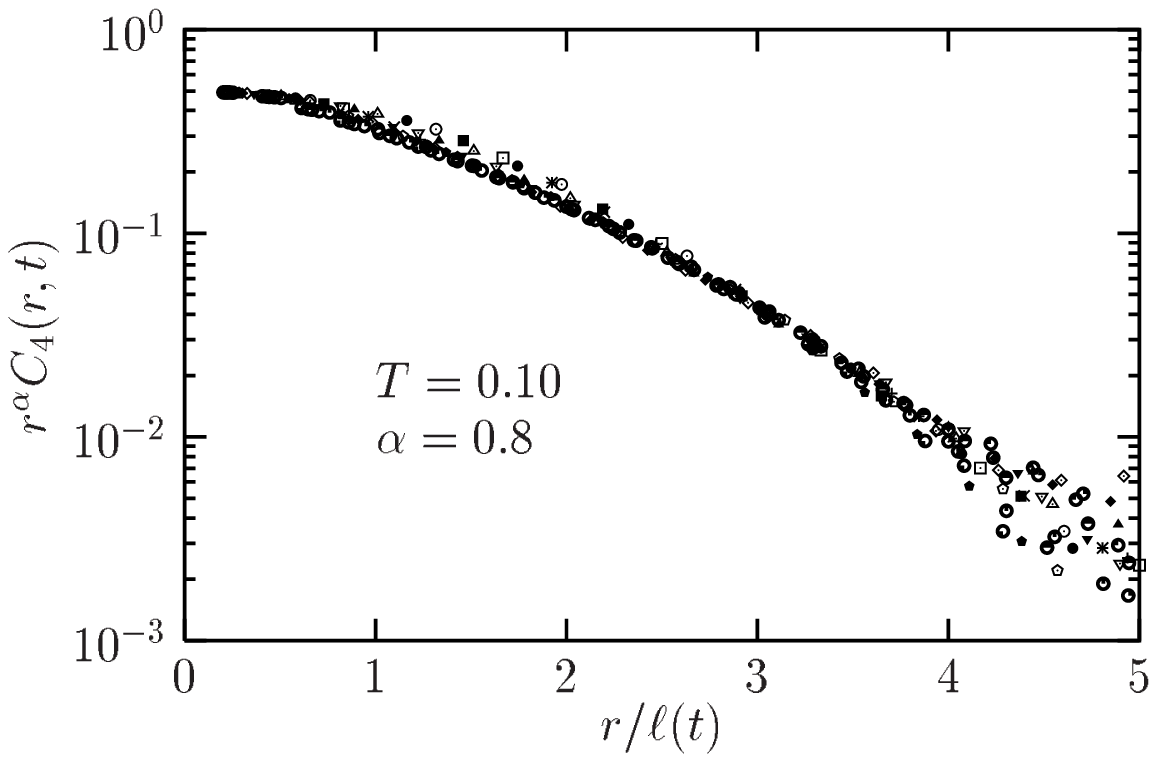,width=8.5cm} 
\psfig{file=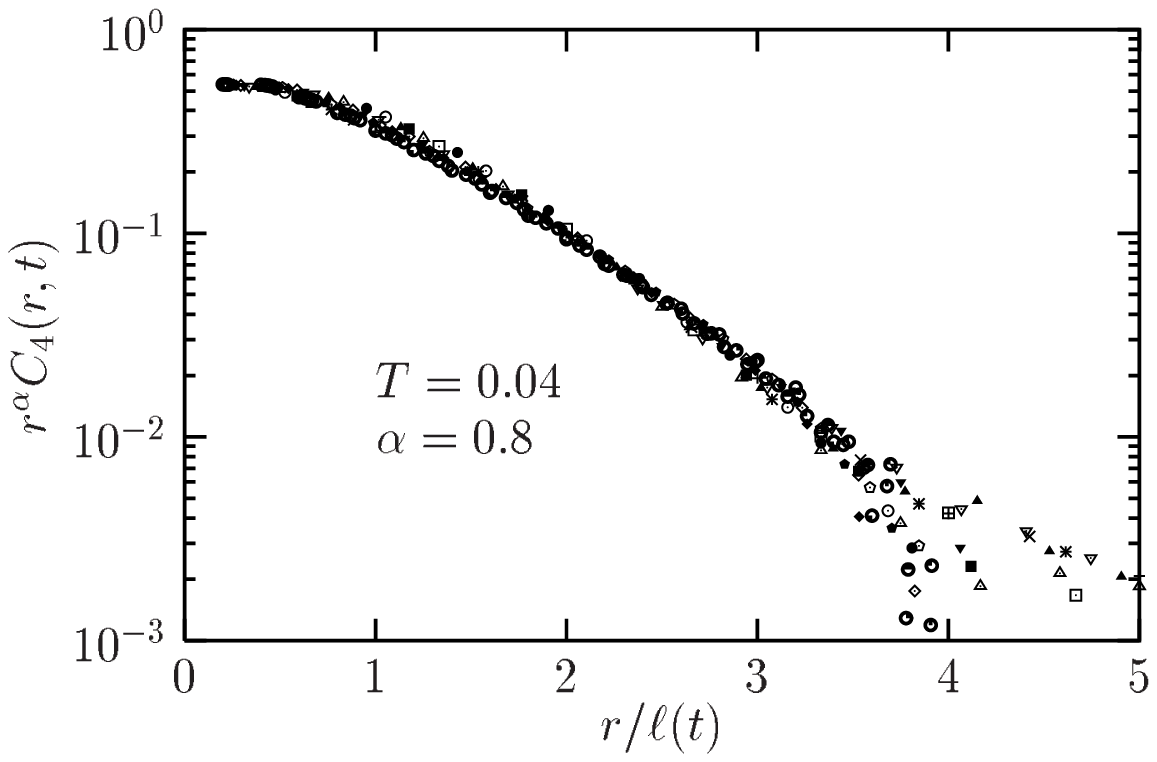,width=8.5cm} 
\caption{\label{scalc4} Rescaled four-point functions 
of the spins according to the scaling form (\ref{c4scal}),
and using (\ref{c4equil}), for temperatures 
$T=0.14$, 0.10 and 0.04 (from top to bottom).
Parameters and symbols are the same as in Fig.~\ref{c4}.}
\end{center}
\end{figure}

We start our scaling analysis with the study of
spatial correlations of the spins.
This choice is dictated by theoretical considerations, since
scaling theories of aging dynamics show that time correlations
have natural scaling forms when expressed as a function of the 
dynamical correlation length 
$\ell$~\cite{fh,encorejp,yosh,surf,yosh5_1,yosh5}, 
so that its knowledge is of primary importance
in this study.

Following the physical discussion of the previous section, it is natural
to suggest the following decomposition of $C_4(r,t_w)$
between a ``locally equilibrated'' and an ``aging'' part,
\be
C_4(r,t_w) \simeq C_{4 {\rm eq}}(r) \, C_{4 {\rm aging}} \left(
\frac{r}{\ell(T,t_w)} \right),
\label{c4scal}
\ee
with $C_{4 {\rm aging}} ( x \to 0) \sim \mbox{constant}$, and 
$C_{4 {\rm aging}} ( x \to \infty) = 0$.
As in studies of the Ising spin glass~\cite{BB,enzo}, we found that
the functional forms
\be
C_{4 {\rm eq}}(r) \simeq r^{-\alpha(T)},
\label{c4equil}
\ee
and
\be
C_{4 {\rm aging}} (x) \simeq \exp \left( - x^\beta \right), \quad \beta > 1,
\ee
represent the data quite well, so that a plot at fixed $T$
of $r^{\alpha} C_4(r,t_w)$ versus the scaling variable $r/\ell$ 
should collapse the data for all times $t_w$. 
Such scaling plots are indeed presented in Fig.~\ref{scalc4}.
Although the data collapse is quite
good, small deviations can be observed in these
scaling plots. This might be due to the fact that
corrections to scaling arise at small waiting times where
the coherence length is still small, so that a scaling
regime defined by $\ell \gg 1$ is not entered yet.
Similar scaling plots were obtained for the Ising spin 
glass~\cite{BB,enzo}, although on a more restricted spatial range.

\begin{table}
\caption{\label{table} Temperature variation
of the exponent $\alpha$ of the power law in Eq.~(\ref{c4equil})
of the spatial correlations, and 
the exponent $\mu$ which occurs in the scaling of 
the autocorrelation function, see Eq.~(\ref{utwt}).}
\begin{ruledtabular}
\begin{tabular}{c c c c c c c c c}
$T$      & 0.16 & 0.15 & 0.14 & 0.12 & 0.10 & 0.08 & 0.04 & 0.02 \\
$\alpha$ & 1.1  & 1.05 & 1.0  & 0.9  & 0.8  & 0.8  & 0.8  & 0.8   \\
$\mu$    & 0.97 & 0.98 & 0.98 & 1.0  & 1.01 & 1.03 & 1.07 & 1.09
\end{tabular}
\end{ruledtabular}
\end{table}

The temperature variation of the exponent $\alpha(T)$ in Eq.~(\ref{c4equil})
is shown in Table \ref{table}.
At very low temperatures, $T \le 0.10$, 
it seems to be roughly constant at about $0.8$.
The same trend is also found in the Ising spin glass, although
there the exponent sticks to the value $0.5$.
At $T_c$ the 
scaling forms in Eqs.~(\ref{c4scal}), and (\ref{c4equil})
are also expected to hold with
$\alpha$ related to the anomalous exponent $\eta$
via the relation
\be 
\alpha (T_c) = d-2+\eta.
\ee
Our estimate for $\eta$ is therefore 
\be
\eta \approx 0.1.
\ee
We cannot estimate
error-bars on this value since it results from 
a somewhat arbitrary scaling procedure. 

As for the Ising case, we find evidence that 
\be
\lim_{r \to \infty} 
\lim_{t_w \to \infty} \lim_{L \to \infty}
C_4(r,t_w) = 0,
\label{obs}
\ee
since $\alpha > 0$. Here $L$ has to be much larger than $\ell(T, t_w)$,
otherwise the system comes to equilibrium, and we get a non-zero value
simply because the system has spin glass order (see again Fig.~\ref{c4L16}).

From the point of view of the droplet
picture\cite{fh}
(which has a single ground state plus those related by global symmetry),
Eq.~(\ref{obs}) is puzzling since one expects 
$\lim_{r,t_w,L \to \infty} C_4(r,t_w) = 
\langle q^2 \rangle_{\rm eq} = q_{\rm EA}^2$.
The data in Figs.~\ref{c4} and \ref{c4chir} are clearly inconsistent
with the large $r$ limits being
equal to $\langle q^2 \rangle_{\rm eq}$ as estimated
from the autocorrelations in Figs.~\ref{auto} and \ref{autochir}.
However, Eq.~(\ref{obs}) has been
justified~\cite{BB,enzo} within the
replica symmetry breaking (many-state, ``RSB'') picture~\cite{BB,enzo}.
The argument~\cite{reviewsimu} 
is that an equilibrium calculation of the correlator
(\ref{c4eq}) within a replica symmetry breaking approach 
predicts power law behavior like Eq.~(\ref{c4equil})
with a non-zero $\alpha$ in the ``zero-overlap'' sector~\cite{cyrano}.
However, this argument is really for equilibrium fluctuations, 
and it is not obvious how to translate this result to the 
non-equilibrium situation of interest here. In particular, 
a restricted average over the ``zero-overlap'' sector
cannot be justified by the sole (trivial) observation 
that the global overlap is zero in the aging regime~\cite{reviewsimu}.

To distinguish between the single and many states pictures, one should perform
local measurements and show that local properties are (in)consistent with the
existence of a single equilibrium state~\cite{fisher}. The correlator in
Eq.~(\ref{c4eq}) can do this, using the (square of the) relative spin
orientation as a local physical observable.  Although,
as discussed above, it is
not clear to us that the argument often given for a power law variation within
RSB theory is correct, the droplet theory makes the clear
prediction~\cite{fh} that
Eq.~(\ref{c4eq}) tends to a constant, $q_{\rm EA}^2$, for $t_w \to \infty$.
Hence the results of our simulations, which fit the power law variation in
Eq.~(\ref{c4equil}) with $\alpha > 0$,
may be closer to the 
RSB ``many states'' picture, at least on the time and length scales being
probed.
Theoretical studies of the two-site, two-replica correlation function, $C_4$,
using the Migdal-Kadanoff approximation
in the spirit of Refs.~\cite{yosh5_1,martin}, 
would be very valuable.

\subsection{Length scales}

\begin{figure}
\begin{center}
\psfig{file=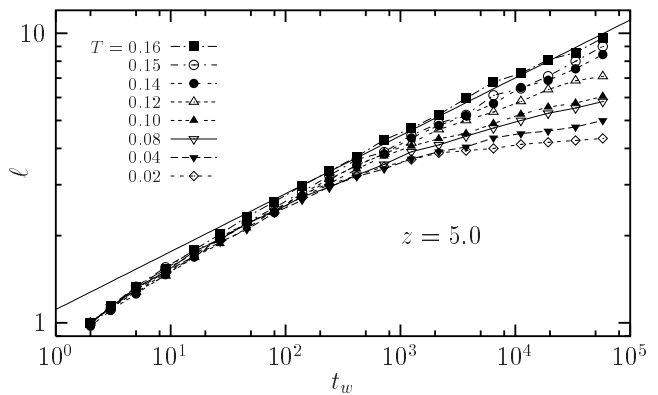,width=8.5cm}
\caption{\label{length}
Growth law of the coherence length at different temperatures.
The data are represented by points and lines.
The straight line is a power law fit to the data at large times for $T=0.16$, 
$\ell \sim t_w^{1/z}$, with $z \approx 5.0$.}
\end{center}
\end{figure}

From the scaling behavior of $C_4(r,t_w)$, we obtain 
a practical definition of the 
dynamical correlation length of the spins
as the quantity leading to the best collapse of the data in Fig.~\ref{scalc4}.
An obvious problem is that the results for $\ell$ 
are affected by the uncertainty in the value of $\alpha$, as is also the 
case in the Ising spin glass. In particular, this makes it impossible
to estimate error-bars on our results for $\ell$.
The evolution with time of the coherence length
at various temperatures is presented in Fig.~\ref{length}, where
a log-log representation is used.
We find that $\ell$ grows with time, 
in a temperature dependent manner, as was anticipated above.

From non-equilibrium critical dynamics arguments, 
one expects $ \ell \sim t_w^{1/z}$, 
where $z$ is the critical dynamic exponent.
An activated scaling, $ \ell \sim (T \ln t_w )^{1/\psi} $,
is more naturally expected at low temperatures, where
$\psi$ is the so-called barrier exponent~\cite{fh}.
Moreover, at a given temperature 
$T \lesssim T_c$, we expect a crossover from a critical (power-law)
regime at short times
to an activated regime at large times, which occurs at shorter times for
lower temperatures. 

All these theoretical expectations can successfully be observed 
in our numerical results in Fig.~\ref{length}.
For $T=0.16$, a power law behavior
is consistent with the data 
at large enough times, with 
\be 
z \approx 5.0.
\ee 
Here again, we did not attempt to define error-bars
on the value of the exponent.
The value $z \approx 7.0$ was found using the same method in the Ising 
spin glass~\cite{BB}.
It is the smaller value of 
the dynamic exponent which allows
larger length scales to be reached in the Heisenberg model than in the Ising
model, as was noted above.

Clearly, at low $T$, the long-time
behavior is not consistent with 
a power law and the bending of the curves is indeed 
compatible with a logarithmic growth law. Note that even at the lowest
temperature we have simulated, $T=0.02$, the first two decades of 
waiting times give a growth law very similar to the one
obtained at $T=0.16$, indicating that the ``true'' 
activated behavior is entered in the last three decades of the simulations
only. This is consistent with the corrections to scaling observed
in the spatial correlators, recall Fig.~\ref{scalc4}.
For this reason, our data do not allow a precise 
determination of the barrier exponent $\psi$ from a logarithmic
fit to the growth law which contains too many free parameters.
Likewise, the interpolation law between critical and
activated regimes~\cite{encorejp,BB,dupuis,yosh6},
\be
t_w \sim t_o \left(\frac{\ell}{\ell_o} \right)^z 
\exp \left[ \frac{\Upsilon_o}{T} 
\left( \frac{\ell}{\xi(T)} 
\right)^\psi \right],
\label{interp}
\ee
which was used in earlier studies,
does not account for the crossover seen in our data.
In this expression, $\xi \sim |T-T_c|^{-\nu}$ where 
$\nu$ is the critical exponent for the 
equilibrium correlation length, and
$\ell_o$, $t_o$ and $\Upsilon_o$ are microscopic 
length, time and energy scales respectively.

The conclusion is that larger time scales need to be studied 
to establish a more quantitative description of the growth law
for $\ell$, but this is difficult because it 
would require a huge amount of computer time, as one would
have to work with even larger system sizes.
 
\begin{figure}
\begin{center}
\psfig{file=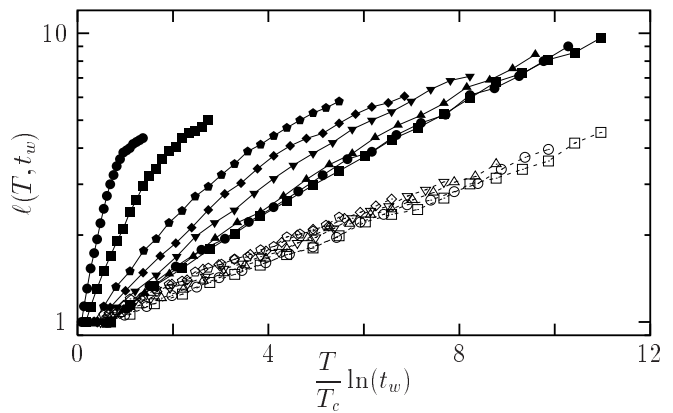,width=8.5cm}
\caption{\label{length3} 
Comparison of the growth of the dynamic correlation
length in Heisenberg (data from Fig.~\ref{length}) 
and Ising (published data from Ref.~\onlinecite{BB}) models.
We use the experimental representation of Ref.~\onlinecite{bert}, 
plotting $\ell$ versus $(T/T_c) \ln(t_w)$.
Heisenberg data are represented with closed symbols
for $T = 0.16$, 0.15, 0.14, 0.12, 0.10, 0.08, 0.04, 0.02 (from 
right to left), $T_c \simeq 0.16$.
Ising data are represented with open symbols
for $T=0.95$, 0.855, 0.76, 0.665, 0.57, 0.475
(from right to left), $T_c \simeq 0.95$.}
\end{center}
\end{figure}

It is nevertheless possible to compare these results with the Ising spin
glass. In Fig.~\ref{length3}, we replot the data of Fig.~\ref{length}
together with published data obtained in the Ising Edwards-Anderson 
model~\cite{BB}.
We use the representation adopted in Ref.~\onlinecite{bert}
to compare similar experimental data in Heisenberg and Ising samples,
and plot 
$\ell$ versus $\frac{T}{T_c} \ln(t_w)$ in a lin-log scale.
In this representation, data at different times and temperatures collapse
if $\ell \sim t_w^{z T/T_c}$, a growth law advocated 
in some early studies~\cite{yosh2,enzo,orbach}. 
We recognize from Fig.~\ref{length3} that this functional form 
indeed reasonably accounts for the Ising data but not at all
for the Heisenberg ones.
Also, Fig.~\ref{length3} clearly confirms
that larger length scales can
be studied in the Heisenberg model than in the Ising model
in three dimensions.
Further, we find in the Heisenberg model
clear evidence of an activated, logarithmic 
growth law, shown by the downward curvature in the data in
Figs.~\ref{length} and \ref{length3} 
at low temperature. This effect is not observed in the Ising model, 
see Fig.~\ref{length3}.

Turning to experiments, direct measurements of the dynamical correlation
length are obviously impossible since the correlator in
Eq.~(\ref{c4eq}) can not be 
measured experimentally. The least indirect method we are aware of
was first suggested in Ref.~\onlinecite{orbach} 
and used more recently in a variety of samples~\cite{bert}.
It consists of a dynamic estimate of the number of spins that relax
in a coherent manner probed through a magnetic perturbation.
Results~\cite{bert} on both Heisenberg-like and Ising-like systems
were gathered in the same representation as Fig.~\ref{length3} (see
Fig.~4 in Ref.~\onlinecite{bert}).
Our results are useful to understand the trends found in 
experiments. 
In particular, in Ising samples, the dynamical correlation length is 
experimentally found to be smaller
but to grow faster with increasing time than in 
Heisenberg samples~\cite{bert}.
We will now see that this can be understood from the data in 
Fig.~\ref{length3}.

The Ising
spin glass has a larger dynamical exponent $z$, which means that
its dynamical correlation length initially grows more slowly, since the
straight-line data at $T=T_c$ in Figs.~\ref{length} and 
\ref{length3} have slope $1/z$.
At intermediate temperatures where experiments are usually carried out,
the system leaves the critical regime after a crossover 
time scale $t_{\rm c}$ given by
\be
t_{\rm c} \sim |T-T_c|^{-z \nu}.
\label{tc}
\ee
For the Gaussian Ising and Heisenberg Edwards-Anderson
models in three dimensions, it is
found that~\cite{BB,peternew}
\be
z \nu |_{\rm Ising} \simeq 10.5, \quad 
z \nu |_{\rm Heisenberg} \simeq 5.5,
\label{znu}
\ee
in fair agreement with the experimental values reported
in Ref.~\onlinecite{bert}:
$10.5$ for the Ising system Fe$_{0.5}$Mn$_{0.5}$TiO$_3$, 
and $7$ and $5$ respectively for the Heisenberg systems
CdCr$_{1.7}$In$_{0.3}$S$_4$ and Ag:Mn $2.7\%$.
Eqs.~(\ref{tc}) and (\ref{znu}) show
that Heisenberg systems leave the critical region 
at much earlier times than Ising samples.
For $t > t_c$ there is a crossover, visible in
Fig.~\ref{length} and
the Heisenberg data in Fig.~\ref{length3},
to presumably activated behavior, $\ell \propto
(\ln t)^{1/\psi}$, where $\psi$ is the barrier exponent.
It can be argued~\cite{BB} that this crossover rationalizes the
apparent $T$-dependent dynamic exponent $zT/T_c$ observed in the
Ising model.
Hence, in the experimental time window, Heisenberg samples 
lie deeper in the activated regime where large length scales 
have been reached but further growth is extremely slow. 
In the range of times available in the simulations, 
we do not see the phenomenon observed experimentally~\cite{bert} that
$\ell$ increases more
slowly for Heisenberg systems than for Ising systems.
However, at longer times, such as those
in the experiments, we expect this would occur since 
the Heisenberg model is clearly 
entering the activated region where
$\ell$ increases only logarithmically with $t$. Hence our data provides a
consistent explanation of the results of Ref.~\onlinecite{bert}.

\subsection{Autocorrelation functions}

Finally, we study the scaling behavior of the spin 
autocorrelation function.
This function has been much studied in Ising spin 
glasses, and has recently been measured in experiments~\cite{herisson} 
which have usually
focused on the more easily accessible thermo-remanent
magnetization~\cite{review1,review1_1}. Both functions are
believed to exhibit similar scaling behavior.

As we did for the spatial correlations, it is natural to decompose 
$C(t+t_w,t_w)$ into an equilibrium and an aging part, 
\be
C(t+t_w,t_w) \simeq C_{\rm eq}(t) \,\, C_{\rm aging} 
\left( \frac{h(t+t_w)}{h(t_w)} \right),
\label{autoscal}
\ee
where $h(t)$ is an unknown function increasing with 
time, and the scaling function $C_{\rm aging}$ has the following
limits: $C_{\rm aging}(x \to 0) = \mbox{constant}$
and $C_{\rm aging}(x \to \infty) = 0$.
Two simple expectations 
for the aging part are the following. (i) $h(t) = t$ leading
to a simple $t/t_w$ scaling of the aging part 
of the data; 
(ii) $h(t) = \ell(t)$, the dynamical correlation length,
which is the basic outcome of 
scaling theories~\cite{fh}, and is also obtained in 
coarsening phenomena. 
It is important to notice that although popular, $t/t_w$ is not
theoretically expected to hold in spin glasses, as already 
emphasized~\cite{BB}. The natural choice
of $h(t) = \ell(t)$ does not lead to $t/t_w$ scaling if the dynamics involves
barrier activation because then $\ell(t)$ grows logarithmically with time,
though $t/t_w$ scaling is obtained for critical dynamics where
$\ell(t) \propto t^{1/z}$. Furthermore,
exactly solvable models have demonstrated the possibility
of more general scaling forms~\cite{reviewth,cuku1,cuku2} that 
we also discuss below.

In analysis of experimental data,
deviations from $t/t_w$ scaling are often described
phenomenologically by the replacement
\begin{equation}
{t \over t_w}\ \to \ t_w^{1-\mu}\, f\left({t \over t_w}\right) \, ,
\label{ttw}
\end{equation}
as the scaling variable in $C_{\rm aging}$,
where $\mu$ is an exponent 
(in practice close to unity) and $f$ a scaling function. The simplest choice
would be $f(x) = x$, but it is more common to take 
\begin{equation}
f(x) = {1 \over 1 - \mu} \left[\left(1 + x\right)^{1-\mu}  - 1\right] \, ,
\end{equation}
so Eq.~(\ref{ttw}) becomes
\begin{equation}
{t \over t_w} \  \to \ 
u(t_w, t) = {1 \over 1-\mu} \left[\left(t +
t_w\right)^{1-\mu}  - t_w^{1-\mu}\right] \, ,
\label{utwt}
\end{equation}
since this attempts to take into account the fact that the real
age of the sample during the measurement is $t+t_w$ rather than $t_w$, see
e.g.~Ref.~\onlinecite{cool}. This function $u(t_w, t)$ corresponds to 
\be
h(t) = \exp \left[ \frac{t^{1-\mu}}{1-\mu}  \right]
\label{hmu}
\ee
in Eq.~(\ref{autoscal}), 
and reduces to $t_w^{1-\mu} (t / t_w)$ for $t \ll t_w$.

\begin{figure}
\begin{center}
\psfig{file=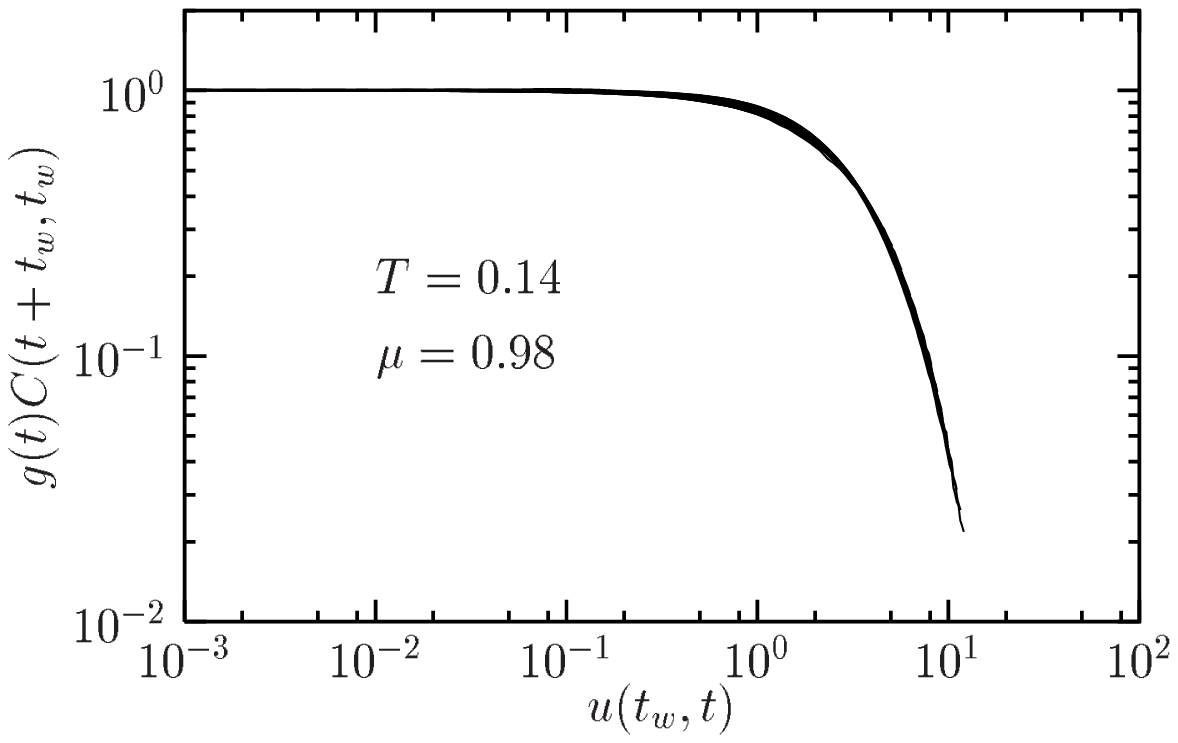,width=8.5cm} 
\psfig{file=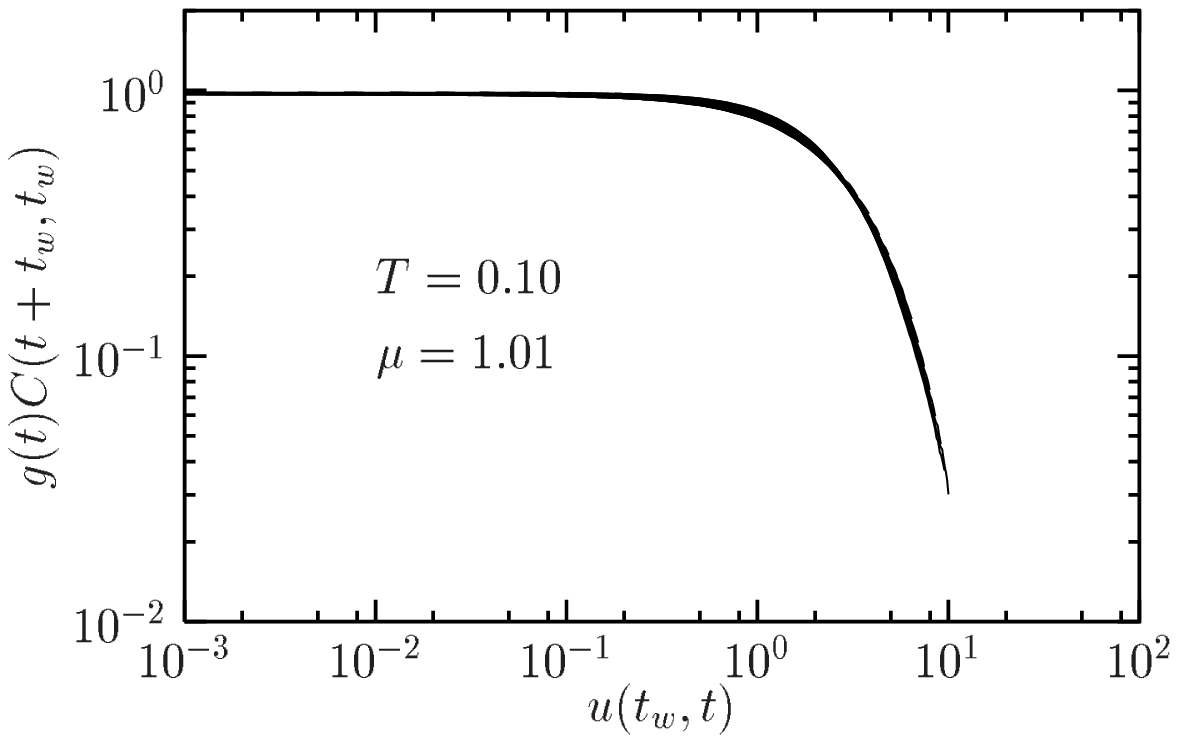,width=8.5cm} 
\psfig{file=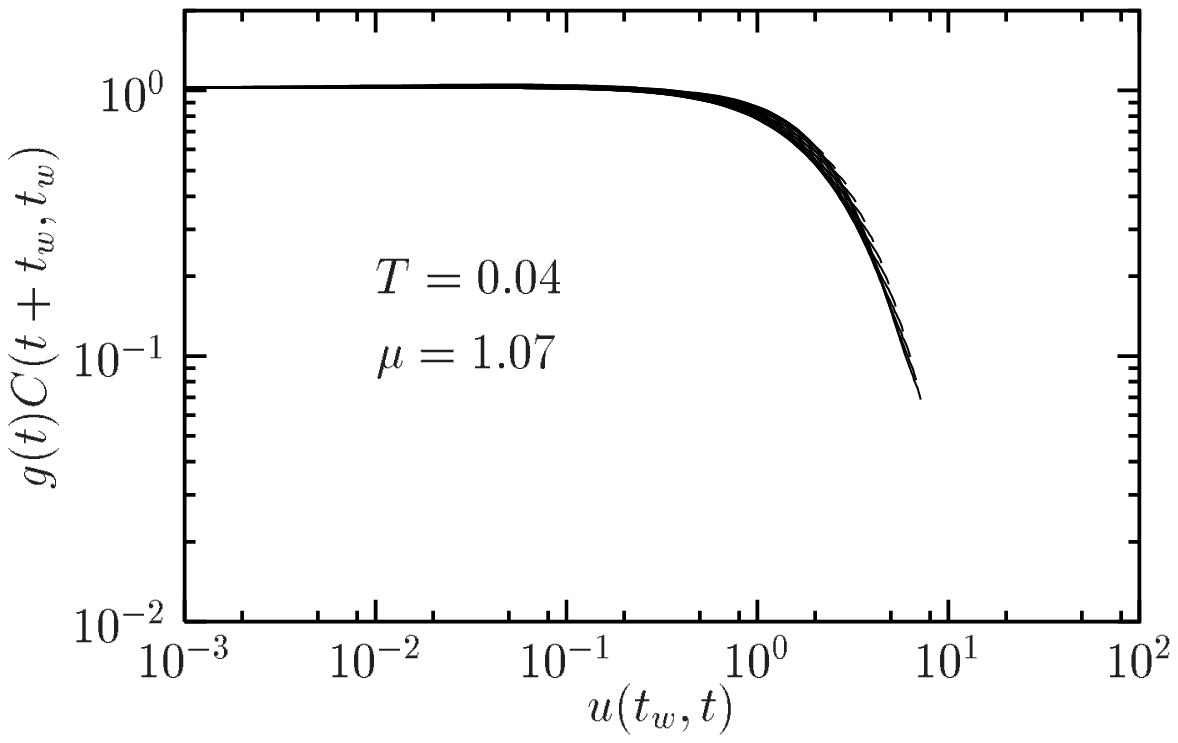,width=8.5cm}
\caption{\label{scalco} Spin autocorrelation
functions rescaled using Eqs.~(\ref{autoscal}), (\ref{utwt}) and 
(\ref{log}). The temperature $T$ and the 
exponent $\mu$ are indicated in each figure.}
\end{center}
\end{figure}

Contrary to the Ising spin glass, we find that the initial
decay of the correlation function is better described
by a logarithmic behavior than by a power law, and we 
take
\be
C_{\rm eq}(t) = a - b \, \log(t) \equiv g(t), 
\label{log}
\ee
where $a$ and $b$ are two temperature dependent parameters
describes the data very well. 
This can be recognized from the initial linear aspect of the data
in Fig.~\ref{auto} where a log-lin representation was used.
Experimentally also, an initial power law decay of the data is 
generally adopted, but the numerical value of this exponent
is extremely small, typically $< 0.10$, so that 
differences from a logarithmic behavior are probably small. 

Although the low temperature data
in Fig.~(\ref{auto}) show evidence for a plateau, i.e. a non-zero value at
large $t_w$ and $t$, Eqs.~(\ref{autoscal}) and
(\ref{log}) do not lead to a true plateau. 
However, Eq.~(\ref{autoscal}) is sufficient to fit the data and has one less
parameter than a function with a plateau.
Also, as discussed many times
in the Ising case~\cite{BB,ricci}, the use 
of an additive scaling in Eq.~(\ref{autoscal}) is in principle
more appropriate, but it too
contains one more free parameter. 
In practice, data taken over a limited 
time window are usually 
not sufficient to independently fix all free parameters
when such an additive scaling is used~\cite{ricci}.  
We find indeed that our data can be equally well fitted with an additive and a 
multiplicative form and both lead to similar results 
as far as the scaling properties of the aging part are concerned.
Here, we shall present results with a multiplicative scaling only. 

With $(a,b,\mu)$ as free parameters, we have been able to 
get a satisfactory collapse of the numerical data at all waiting
times and temperatures by plotting 
$g(t) \, C(t+t_w,t_w)$ as a function of the scaling
variable $u(t_w, t)$.
Representative results are shown in 
Fig.~\ref{scalco}.

We find that the scaling function $C_{\rm aging}(x)$ behaves 
at large $x$ as
\begin{eqnarray}
& C_{\rm aging}(x)  \sim x^{-\lambda_1}, \quad & T \lesssim T_c, \\
& C_{\rm aging}(x)  \sim ( \log x )^{-\lambda_2}, \quad & T \ll T_c, 
\end{eqnarray}
with two temperature dependent 
exponents $\lambda_1, \lambda_2 > 0$.
We find for instance that $\lambda_1 (T=0.14) \approx 0.5$,
$\lambda_1(T=0.12)=0.4$, $\lambda_2 (T=0.04) \approx 2$, 
and $\lambda_2 (T=0.02) \approx 1.5$. 
The existence of two different scaling forms is another 
evidence that the dynamics crosses over from a critical
to an activated regime when $T$ is decreased from $T_c$ down to $0$.
Again, this is not the case in the 
Ising spin glass which is well described by a power law
of the autocorrelation at large times in the whole
low temperature phase, $T \le T_c$~\cite{rieger}.

An outcome of the scaling procedure 
of Fig.~\ref{scalco} is therefore 
an estimate of the temperature dependence
of the exponent $\mu$, and we report this in Table \ref{table}. 
We observe a systematic increase of $\mu$ when the temperature decreases.
For $T=T_c$, we find $\mu < 1$, 
a behavior which has been called ``subaging''. When decreasing the 
temperature, however, $\mu$ becomes  equal to 1 around $T \approx 0.12$, 
and ``superaging'', corresponding to $\mu > 1$,
is found at still lower temperatures.

In the Ising spin glass, the situation is again different since data
in three dimensions indicate that the scaling variable 
$t/t_w$ can be used, corresponding to $\mu=1$,
in the whole low temperature phase~\cite{BB,jimenez}. 
In four dimensions, however, the same tendency as here is found with 
$\mu =1$ close to $T_c$ but $\mu > 1$ at lower temperatures~\cite{BB}. 
More puzzling is the experimental situation, since there one has
$\mu < 1$ and the temperature dependence is the opposite
with $\mu$ decreasing when 
the temperature is lowered~\cite{review1}. 
As discussed in Refs.~\onlinecite{BB,cool}, 
however, experimental quenches 
involve a finite cooling rate, so that laboratory aging
experiments are in fact temperature-shift protocols.
We will therefore come back to subaging behavior 
found in experiments in our future paper~\cite{prep}.

This multiplicity of experimental and numerical behaviors
obviously requires some discussion. 
As described above, there is no obvious theoretical reason
to expect a perfect $t/t_w$ scaling in spin glasses aging
at low temperatures, so that
deviations from this simple behavior should not come as a surprise. 
Moreover, the rescaling obtained in Fig.~\ref{scalco}
is purely phenomenological, and must be interpreted as 
an effective description of the data rather than a fundamental one. 
This is consistent with Kurchan's work showing that persistent 
superaging encoded by Eqs.~(\ref{autoscal}) and (\ref{hmu}) with $\mu >1$
is strictly impossible~\cite{jorge}.

For $T=T_c$, however, critical scaling with $\ell \sim t^{1/z}$ 
naturally implies the use of $\ell(t+t_w)/\ell(t_w)$ and therefore $t/t_w$ 
as scaling variables.
Hence, it is somewhat surprising that the numerics indicate 
instead $\mu(T=0.16) = 0.98$. Since $\mu \approx 1$ is found at a slightly
lower temperature, a possible interpretation could be 
that the value of $T_c$ determined in Ref.~\onlinecite{peternew} 
is slightly larger
than the real one. An indication that this is possible stems
from a recent work~\cite{ian2} where it is argued that the finite
size scaling analysis of Ref.~\onlinecite{peternew} 
systematically overestimates the
critical temperature. A second plausible explanation 
is that our data are plagued by unknown corrections to scaling
in our limited time window at $T=0.16$. 

\begin{figure}
\begin{center}
\psfig{file=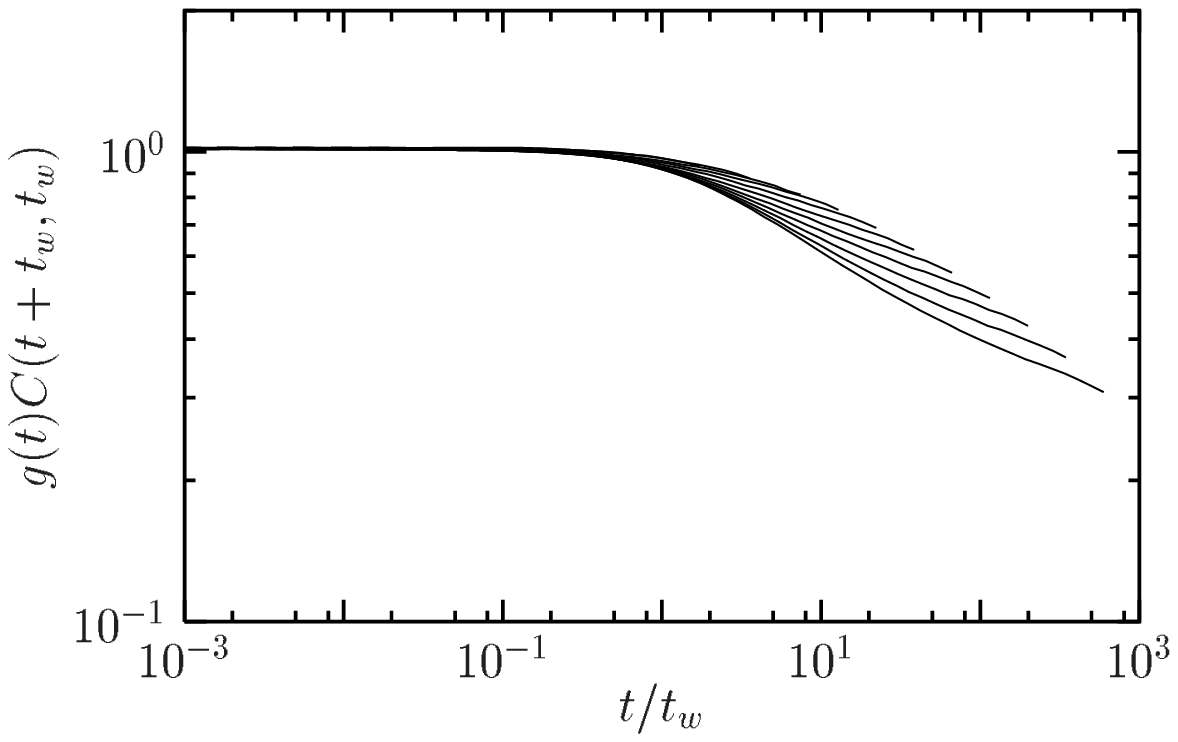,width=8.5cm} 
\psfig{file=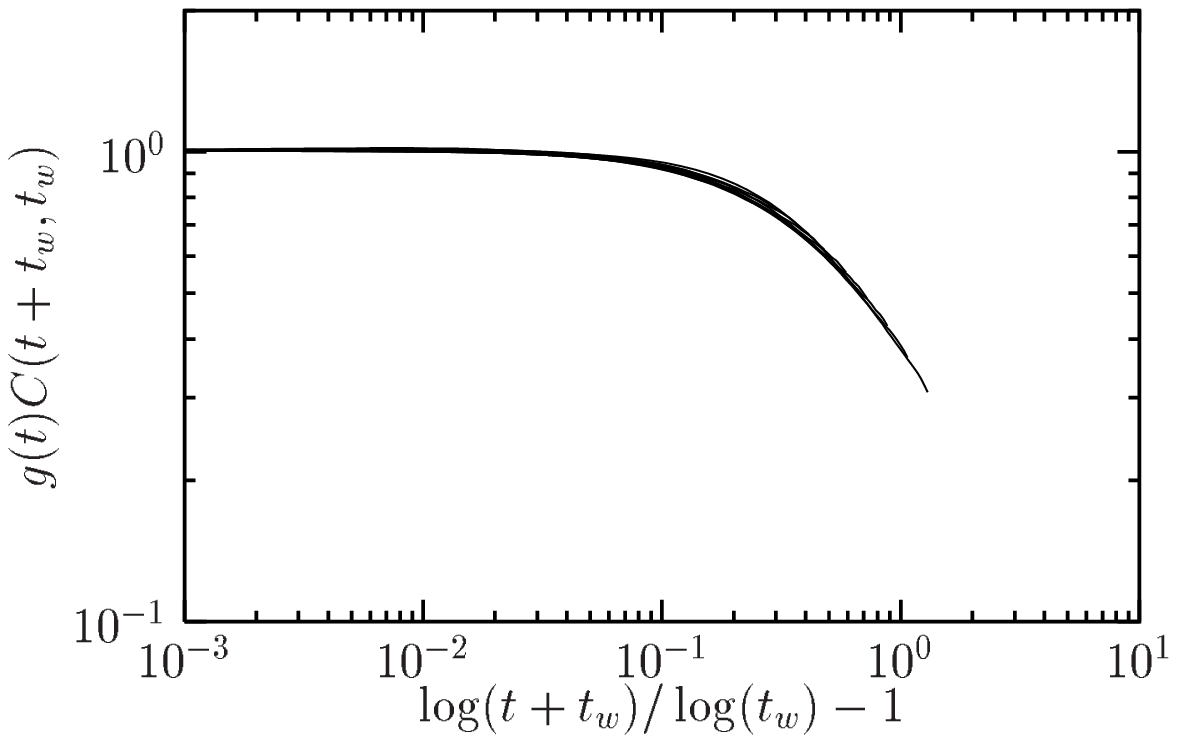,width=8.5cm} 
\psfig{file=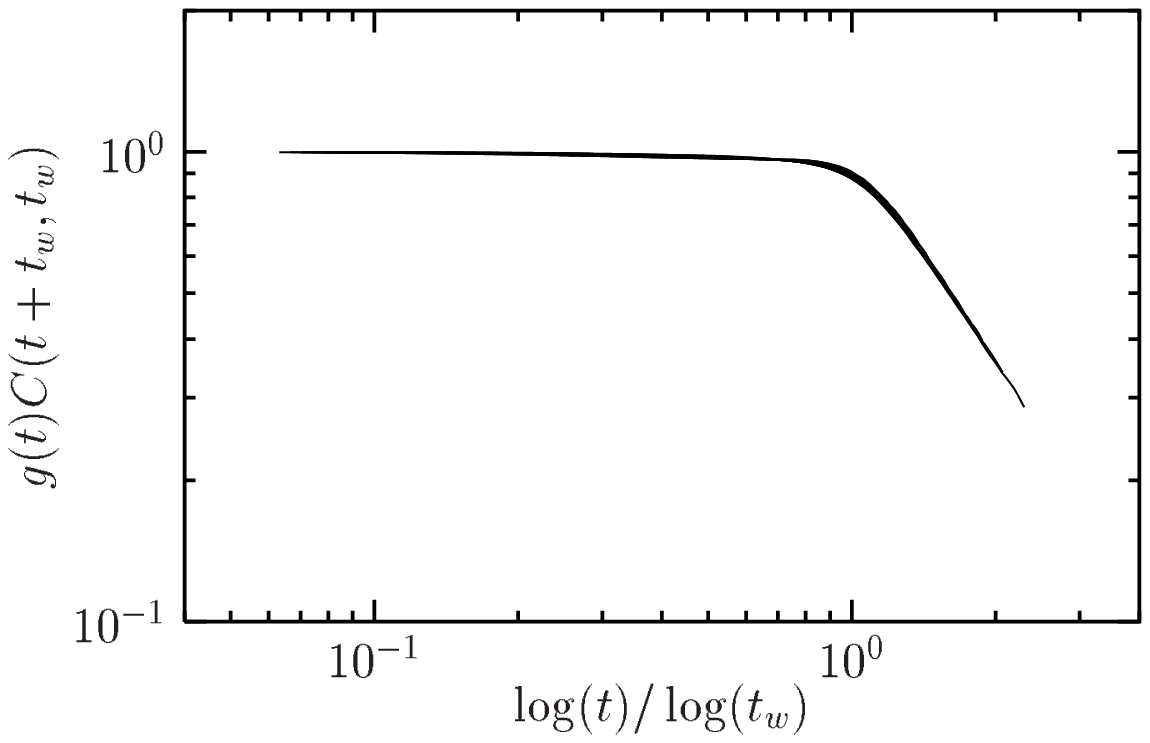,width=8.5cm}
\caption{\label{scalco04} Spin autocorrelation
functions at $T=0.04$ rescaled using 
$t/t_w$, $\log(t+t_w) / \log(t_w)$ or 
$\log(t) / \log(t_w)$ as scaling variables 
for the aging part (from top to bottom).
We only display the 12 largest waiting times, i.e. 
$t_w \in [139, 57797]$.}
\end{center}
\end{figure}

An increasing $\mu$ when temperature is decreased can be rationalized 
as follows.
It is convenient to define an effective relaxation time~\cite{review1_1,BB}
by
$h(t_w)/h'(t_w)$, see Eq.~(\ref{autoscal}). When $h(t) \sim \log(t)$ this
becomes $t_w \log(t_w)$, which is greater than $t_w$,
leading to an effective superaging
behavior since the effective relaxation time grows faster than $t_w$.
The conclusion is that
a crossover towards activated behavior at low $T$ naturally
implies an apparent superaging behavior well consistent
with our numerics. This argument was already used in 
Ref.~\onlinecite{BB} 
to interpret data in the four dimensional Ising spin glass.

A purely activated behavior together with a simple scaling
of the aging part of the decorrelation would imply that 
the use of $\log (t+t_w) / \log(t_w)$ can rescale the data
at large times~\cite{fh}. We test this idea in the middle 
frame in Fig.~\ref{scalco04}.
A comparison with the top frame where $t/t_w$ is used
shows that a logarithmic rescaling of the data
is much better at large times.
To our knowledge, this is the first time that a pure logarithmic rescaling of 
two-time quantities has successfully been employed in a spin glass
simulation. 
This result is important because one of the salient predictions
of the scaling approach to spin glasses is the prediction of
activated dynamics and logarithmic scaling in the aging regime
of finite dimensional spin glasses~\cite{fh}.
Experimentally, a logarithmic rescaling of the thermo-remanent
magnetization does not collapse the data~\cite{prep,these}. It does, 
however, rescale 
data for the out-of-phase susceptibility
at different frequencies $\omega$~\cite{yosh4,these},
but in a much more restricted time regime, $\omega t_w \gg 1$.

Theoretical insights are also provided by the exact solution 
of the non-equilibrium dynamics of mean-field spin glasses
which has revealed the existence of dynamic 
ultrametricity~\cite{cuku1,cuku2,BBK}. 
Physically, ultrametricity implies the existence
of a broad distribution of relaxation times 
organized in a hierarchical manner, in the very precise sense 
described in Ref.~\onlinecite{cuku1}. 
Technically, this generalizes Eq.~(\ref{autoscal})
to a continuum of scaling functions $C_{\rm aging}$ associated 
with a continuum of functions $h(t)$.
Interestingly, dynamic ultrametricity also arises in another 
mean-field solvable model which has the advantage that
an exact form for the time decay can be worked out~\cite{eric}. 
There it was found 
that two-time correlators scale with the variable $\log(t) / \log(t_w)$.
This is subtly different from the logarithmic rescaling
used above but the two scaling variables are not compatible, and only 
the latter leads to dynamic ultrametricity. We have 
applied this alternative rescaling in the lower frame 
in Fig.~\ref{scalco04}, which
shows that it is marginally superior to the logarithmic
rescaling used above. 
Note however that this alternative scaling variable 
``compresses'' slightly more the numerical data, 
so that we regard both scalings as being of equally good quality. 
Note too that they are both compatible,
in a restricted time window, with an effective description
in terms of a superaging behavior.
This is again the first time, to our knowledge, that 
aging data so directly 
hint at the possible existence of dynamic ultrametricity
in a three-dimensional spin glass. 

\section{Summary and conclusion}
\label{conclusion}

We have performed the first large scale
numerical simulations of the
the aging dynamics of the 
three-dimensional Heisenberg spin glass. We have measured and
discussed in detail the behavior of several space-time 
correlators related both to spin and chiral degrees of freedom. 
We now summarize our main results.

We find that the non-equilibrium dynamics of spins and chiralities 
are qualitatively very similar in the range $T \in [0.02, 0.16]$
that we have investigated in detail. 
Two-time autocorrelation functions show the existence
of non-zero Edwards-Anderson parameters
for spins and chiralities, and the large time 
behavior is qualitatively similar and exhibit aging
behavior, $C(t+t_w,t_w) \ne C(t)$, indicating the 
simultaneous freezing of
both spins and chiralities in this temperature window.

Appropriate spatial correlation functions show 
that aging is due to the development with time of a spin dynamical correlation
length, 
so that aging in spin glasses can be thought of as a sort of coarsening, as 
proposed long ago by Fisher and Huse~\cite{fh}. 
All our results are consistent with the growth with time of a random 
ordering of the spins imposed by the quenched disorder. 
The chiralities naturally follow the spins but display weaker
correlations. This is because spins correlations
fall off with power of distance in the non-equilibrium regime, so that 
the correlations of the
chiralities, each of which involves three spins, 
probably fall off with a faster power law.

The system's behavior
can be simply accounted for without invoking 
any decoupling between spins and chiralities,
and we have explicitly shown that 
some of the numerical evidence apparently supporting
such a decoupling suffers from strong finite size effects. 

We find that the growth law of the spin dynamical correlation
length 
changes from a critical power law regime for temperatures close
to $T_c$ to a slower activated regime at lower temperatures.
This crossover naturally influences the scaling behavior
of two-time quantities, and we have found evidence 
that the low temperature behavior of two-time autocorrelations
is more naturally interpreted in terms of 
logarithmic rescalings than the standard $t/t_w$ scaling
commonly used. In particular, predictions from both droplet 
or mean-field approaches to spin glass dynamics are equally able to 
rationalize the time scaling of dynamical functions.

Although qualitatively similar at first sight, we have found 
that the Heisenberg Edwards-Anderson model in three dimensions
behaves quantitatively
differently from its Ising cousin in several ways:
\begin{enumerate}
\item
The Heisenberg model clearly displays an Edwards-Anderson
parameter in the dynamics, see Eq.~(\ref{ea}) and Fig.~\ref{auto}, whereas the
Ising model does not.
\item
The growth of the coherence length is not described 
by a power law but exhibits a rapid crossover towards 
activated dynamics, see Figs.~\ref{length} and \ref{length3}.
\item
Crossover to activated dynamics is 
also observed in the scaling 
of two-time quantities which are not 
described by $t/t_w$ scaling in the low temperature phase, see
Fig.~\ref{scalco04},
\item
Much larger length scales are involved in the dynamics
at large times and low temperatures, due to the smaller value of the 
product $z \nu$ of critical exponents, see Eq.~(\ref{znu}).

\end{enumerate}

Interestingly, the features listed above that distinguish the Heisenberg
from the Ising Edwards-Anderson model also make 
the aging dynamics of the Heisenberg model much closer to the experimental
situation where results are generally more naturally interpreted  
in terms of thermally activated dynamics and logarithmically 
slow relaxations~\cite{review1,review1_1}. 
As far as simple aging protocols are concerned, we have clearly 
established that aging behavior has to be interpreted in
terms of the slow growth with time of a spin glass dynamic
correlation length which then dictates scaling behavior of physical
properties. Although direct experimental
access to $\ell(t_w)$ is hard~\cite{orbach}, 
recent comparative studies~\cite{bert}
of Ising and Heisenberg samples are in full agreement with our findings that
Ising samples display faster growth of a smaller dynamic correlation length
on the time scales of interest, see Fig.~\ref{length3}.
However, despite these similarities, we have noted that the logarithmic
scalings found in this work 
at low temperatures and large times are not 
necessarily observed in
experiments because of the finite cooling rates used in 
experiments~\cite{BB,cool}.
Finally, our observation that the Heisenberg model, as opposed 
to the Ising version, displays a clear crossover towards
activated dynamics is in qualitative agreement with a series of
recent studies~\cite{dupuis,yosh3,bert} showing that memory effects 
and the influence of temperature shifts
are much stronger in Heisenberg samples. 

These conclusions, as well as our preliminary results~\cite{prep}, 
point to the correctness of the initial intuition
which motivated this work that the present model
is better-suited to study and understand from
a microscopic viewpoint also more complex thermal protocols
leading to further non-equilibrium effects. 

\begin{acknowledgments}
We thank J.-P. Bouchaud, I. Campbell, E. Vincent for discussions, 
and A. Dupuis for his kind help during the preparation of the manuscript.
The work of LB is supported by the EU through a 
Marie Curie Grant No.\ HPMF-CT-2002-01927, CNRS France, 
and Worcester College Oxford. The work of APY is supported by the NSF through
grants DMR 0086287 and 0337049.
The simulations were performed at the Oxford Supercomputing 
Center, Oxford University.
\end{acknowledgments}

\end{document}